\documentclass[
    reprint,
    aps,
    prl,
    superscriptaddress,
]{revtex4-2}

\usepackage[version=4]{mhchem}
\usepackage{amsmath}
\usepackage{amssymb}
\usepackage{booktabs}
\usepackage{glossaries}
\usepackage{graphicx}
\usepackage{tabularx}
\usepackage{siunitx}
\usepackage{hyperref}

\graphicspath{{./figures/}}

\hypersetup{
    pdfauthor={},
    pdftitle={},
    pdffitwindow=false,
    pdfstartview={FitH},
    pdfnewwindow=true,
    colorlinks=true,
    linkcolor=black,
    citecolor=black,
    filecolor=black,
    urlcolor=black,
}

\setacronymstyle{long-short}
\newacronym{dft}{DFT}{density functional theory}
\newacronym{md}{MD}{molecular dynamics}
\newacronym{nep}{NEP}{neuroevolution potential}
\newacronym{rp}{RP}{Ruddlesden-Popper}
\newacronym{xrd}{XRD}{X-ray diffraction}

\newcommand{\hmn}[1]{
  \ensuremath{\begingroup\setupHMN #1\endgroup}%
}
\newcommand{\setupHMN}{%
  \doHMN{-}{\HMNoverline}%
  \doHMN{*}{\HMNminverse}%
  \doHMN{i}{\infty}
}
\newcommand{\doHMN}[2]{%
  \begingroup\lccode`~=`#1
  \lowercase{\endgroup\let~}#2%
  \mathcode`#1="8000
}
\newcommand{\HMNminverse}[1]{\frac{#1}{m}}
\newcommand{\HMNoverline}[1]{\mkern1mu\overline{\mkern-1mu#1\mkern-1mu}\mkern1mu}

\DeclareSIUnit\angstrom{\text{Å}}
\DeclareSIUnit\atom{\text{atom}}

\newcommand{\chalmers}{Department of Physics, Chalmers University of Technology, SE-41296, Gothenburg, Sweden}
\newcommand{\toon}{School of Engineering, Physics and Mathematics, Northumbria University, Newcastle upon Tyne, NE1 8QH, United Kingdom}

\begin{document}


\title{\textbf{Diverse polymorphism in Ruddlesden-Popper chalcogenides} 
}

\author{Prakriti Kayastha}
\affiliation{\toon}
\author{Erik Fransson}
\affiliation{\chalmers}
\author{Paul Erhart}
\affiliation{\chalmers}
\author{Lucy Whalley}
\email{l.whalley@northumbria.ac.uk}
\affiliation{\toon}

\date{\today}

\begin{abstract}
\Gls{rp} chalcogenides are an emerging class of layered semiconductors with tunable properties and chemical stability, making them promising candidates for a wide range of functional applications.
Over the past four decades, the structural diversity of \gls{rp} oxides has been exploited to realise advanced functionalities; however, similar strategies have not yet been applied to \gls{rp} chalcogenides, whose structural behaviour remains poorly understood.
In this study, we develop a high-accuracy machine-learned interatomic potential to perform large-scale molecular dynamics simulations of the homologous \gls{rp} series \ce{Ba_{n+1}Zr_nS_{3n+1}}.
We identify new polymorphs for each $n$-value, predict the corresponding phase transition temperatures, and validate our approach through comparison with existing experimental data. 
We find that the $n=1$ phase exhibits in-plane negative thermal expansion, that the $n=1$ and $n=3$ phases undergo unusual ascending symmetry breaking, and that phases with $n\geq3$ develop layer-dependent tilt patterns not previously observed in inorganic \gls{rp} compounds.
This distinctive behaviour arises from the interplay between \ce{ZrS6} octahedral rotations and \ce{BaS} rumpling at the perovskite-rocksalt interface, suggesting new strategies for realising advanced functionalities and tuning properties in \gls{rp} chalcogenides.
\end{abstract}

\maketitle

Three-dimensional networks of corner-sharing octahedra constitute the fundamental structural units of \ce{ABX3} perovskite materials.
Layered perovskites, which consist of single or multiple perovskite layers separated by thin inorganic or organic spacers, have also been widely investigated.
Perovskite oxides and their layered counterparts have received extensive attention due to their pronounced structural versatility, which can be exploited to probe fundamental physical phenomena and enable advanced functionalities such as superconductivity, ferroelectricity, and negative thermal expansion \cite{Bednorz1988perovskite,Benedek2015understanding,McCabe2015proper,Ablitt2017}.
A crucial step toward designing new perovskite-based functional materials is thus identifying the accessible polymorphs; however, this remains challenging because the structural distinctions between polymorphs are often subtle and difficult to resolve experimentally or theoretically \cite{Woodward2005Electron}.

\begin{figure}[hb!]
    \centering
    \includegraphics[width=\linewidth]{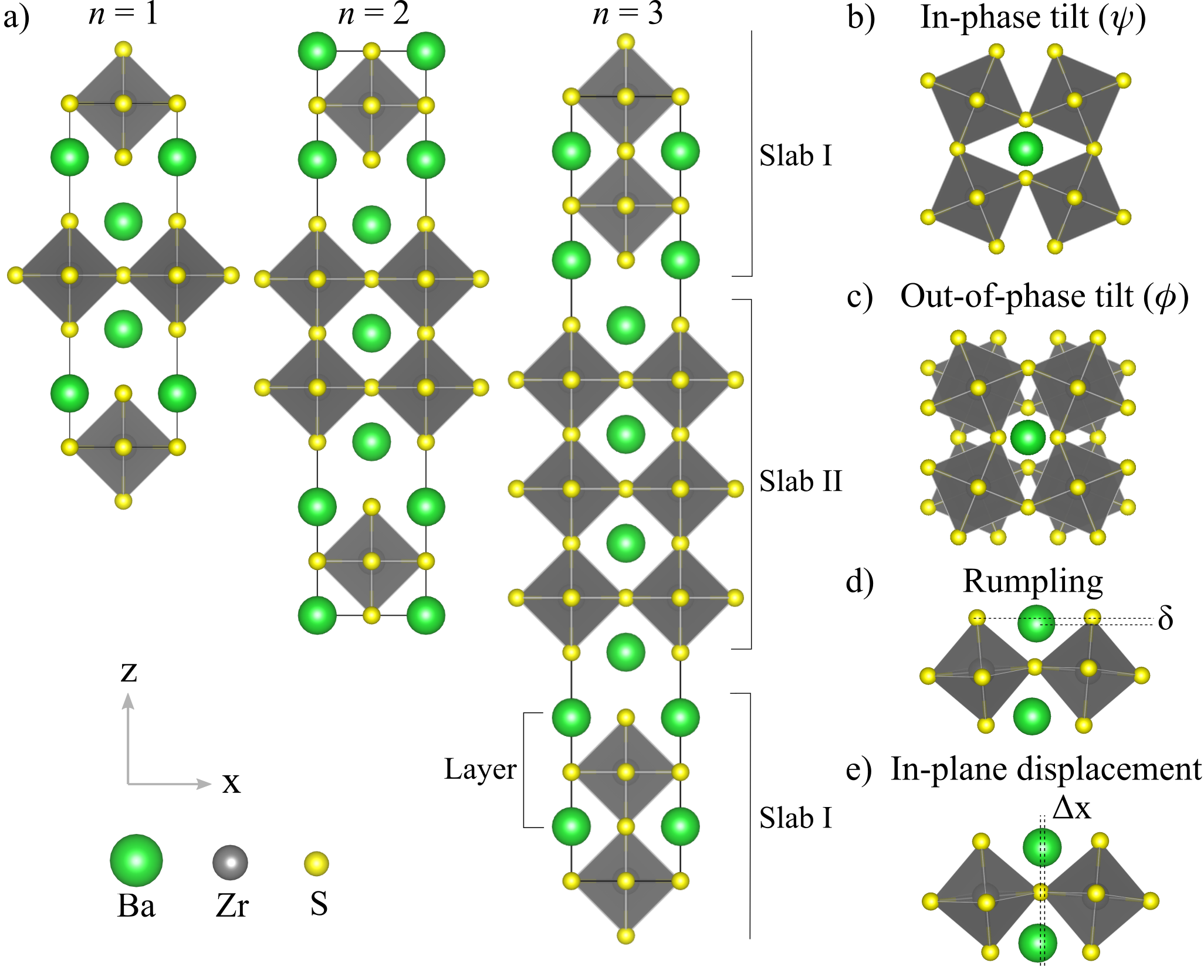}
    \caption{
    a) \ce{Ba_{n+1}Zr_{n}S_{3n+1}} Ruddlesden-Popper (RP) crystal structures in the high-symmetry \hmn{I4/mmm} phase.
    The primitive cell is formed from two unique perovskite slabs with a half-cell offset.
    Each slab contains $n$ layers of \ce{ZrS6} octahedra; the $n=\infty$ limit is equivalent to \ce{BaZrS3}.
    Common distortions in RP materials: b) in-phase ($\psi$) and c) out-of-phase ($\phi$) octahedral tilting; d) out-of-plane rumpling distortion; e) in-plane cation displacement.
    The rumpling amplitude is defined as positive when A-site cations at the rocksalt layer move towards the \ce{BX6} octahedra.
    In-plane refers to displacements along the $x$- or $y$-axes.}
    \label{fig:crystal_structures_phonons}
\end{figure}

Perovskite-based chalcogenides (X = S, Se) are an emerging class of tunable semiconductors for optoelectronic and thermoelectric applications \cite{sopiha2022chalcogenide, jaramillo2019praise, niu2018optimal, kayastha2023high, osei2021examining, wu2023ultralow, hoque2025ruddlesden}. 
Among them, \ce{BaZrS3} is the most studied, exhibiting strong visible-light absorption and photoluminescence, defect tolerance, and chemical stability \cite{nishigaki2020extraordinary, nielsen2025lights, wu2021defect, yuan2024assessing}. 
The \ce{Ba_{n+1}Zr_nS_{3n+1}} series is known to form in the \gls{rp} structure, in which perovskite slabs are separated by rocksalt layers \cite{chen1993preparation, hung1997ba3zr2s7, niu2018thermal, pradhan2024emergence,saeki1991preparation, chen1994structural} (\autoref{fig:crystal_structures_phonons}a).
\ce{Ba2ZrS4} ($n=1$) is reported in the high-symmetry \hmn{I4/mmm} phase with untilted octahedra, \ce{Ba3Zr2S7} ($n=2$) is reported in both the \hmn{I4/mmm} and \hmn{P4_2/mnm} phases, and \ce{Ba4Zr3S10} ($n=3$) in the \hmn{Fmmm} phase.
While distortions to other phases are expected upon cooling or heating, as in other perovskite-like materials, these transitions remain largely unexplored in \gls{rp} chalcogenides, leaving a critical gap in understanding their structural evolution and the potential for novel functional properties.

Here, we address this gap by combining a high-accuracy machine-learned interatomic potential with large-scale molecular dynamics, revealing a rich polymorphic landscape and unexpected structural behaviour across $n=1$–6. 
This approach identifies the onset of 3D perovskite-like behaviour and shows how structural distortions evolve with temperature and $n$-value.

For each $n$-value, we performed \gls{dft} calculations of energies, forces, and stresses for the 33 unique tilt patterns enumerated by Aleksandrov and Bartolomé \cite{blum2009ab,knoop2020fhi,aleksandrov2001structural}. 
This data formed the initial training set for a \gls{nep} model, which enables efficient molecular dynamics sampling of the potential energy surface \cite{FanZenZha21, Fan22, FanWanYin22}. 
The final model, trained on 1375 \gls{rp} and perovskite HSE06-evaluated structures, achieves a formation energy root mean squared error of \qty{1.8}{\milli\electronvolt\per\atom}, providing a highly accurate representation of the energy landscape \cite{heyd2003hybrid}.
Additional computational details, including a comparison of \gls{dft}- and \gls{nep}-calculated phonon spectra, are provided in the Supplemental Material \cite{supp_mat,Larsen2017,calorine,perdew2008restoring,pacsca2025machine,momma2011vesta,krukau2006influence,FanWeiVie2017,Togo2024,stukowski2009visualization,EriFraErh19,fransson2023phase,fransson2024impact,FraRosEriRahTadErh23}.

\begin{table}
\caption{
    Predicted ground state tilt patterns and resulting space groups for \ce{Ba_{n+1}Zr_nS_{3n+1}} materials, where $n$ is the number of perovskite layers within each slab.
    Aleksandrov notation is used to denote the out-of-phase ($\phi$) and in-phase ($\psi$) octahedral tilts along each axis, with an overbar to indicate that the tilt patterns between adjacent slabs are in opposite directions \cite{aleksandrov1987successive}.
}
\label{tab:ground_state}
\begin{tabular}{llll}
\toprule
 $n$ & Slab I & Slab II & Space group  \\ 
\midrule
 1& $\phi$00 &  0$\Bar{\phi}$0  & \hmn{P4_2/ncm} \\
 2& $\phi$00  & 0$\phi$0 & \hmn{P4_2/mnm} \\
 3& $\phi$00  & 0$\Bar{\phi}$0 & \hmn{P4_2/ncm}  \\
 4&  $\phi\phi\psi_z$ & $\phi\phi\Bar{\psi}_z$ & \hmn{Pnma}  \\
 5&   $\phi\phi\psi_z$ &  $\Bar{\phi}\Bar{\phi}\Bar{\psi}_z$   & \hmn{P2_1/c}    \\
 6&  $\phi\phi\psi_z$ & $\phi\phi\Bar{\psi}_z$  & \hmn{Pnma}   \\
\bottomrule
\end{tabular}
\end{table}

\autoref{tab:ground_state} shows the predicted ground state tilt patterns of the \gls{rp} series, which can be divided into two groups. 
Low-$n$ phases ($n = 1$–3) feature a single out-of-phase tilt along the $x$-axis in slab I and along the $y$-axis in slab II. 
High-$n$ phases ($n = 4$–6) exhibit enhanced tilting in both slabs, with out-of-phase tilts along the $x$- and $y$-axes, and an in-phase tilt along the $z$-axis.
\ce{BaZrS3} adopts the same tilt pattern in its low-temperature \hmn{Pnma} structure, indicating that for $n \ge 4$, the \ce{Ba_{n+1}Zr_nS_{3n+1}} ground states converge to that of the parent perovskite. 
Across all $n$-values, the ferroelectric \hmn{Cmc2_1} phase is metastable but lies very close in energy, within \SI{2}{\milli\electronvolt\per\atom}, to the paraelectric ground state. \gls{nep}-calculated energies for all 198 tilt patterns are provided in Tables S1 and S2.

\begin{figure}
\centering
\includegraphics[width=\linewidth]{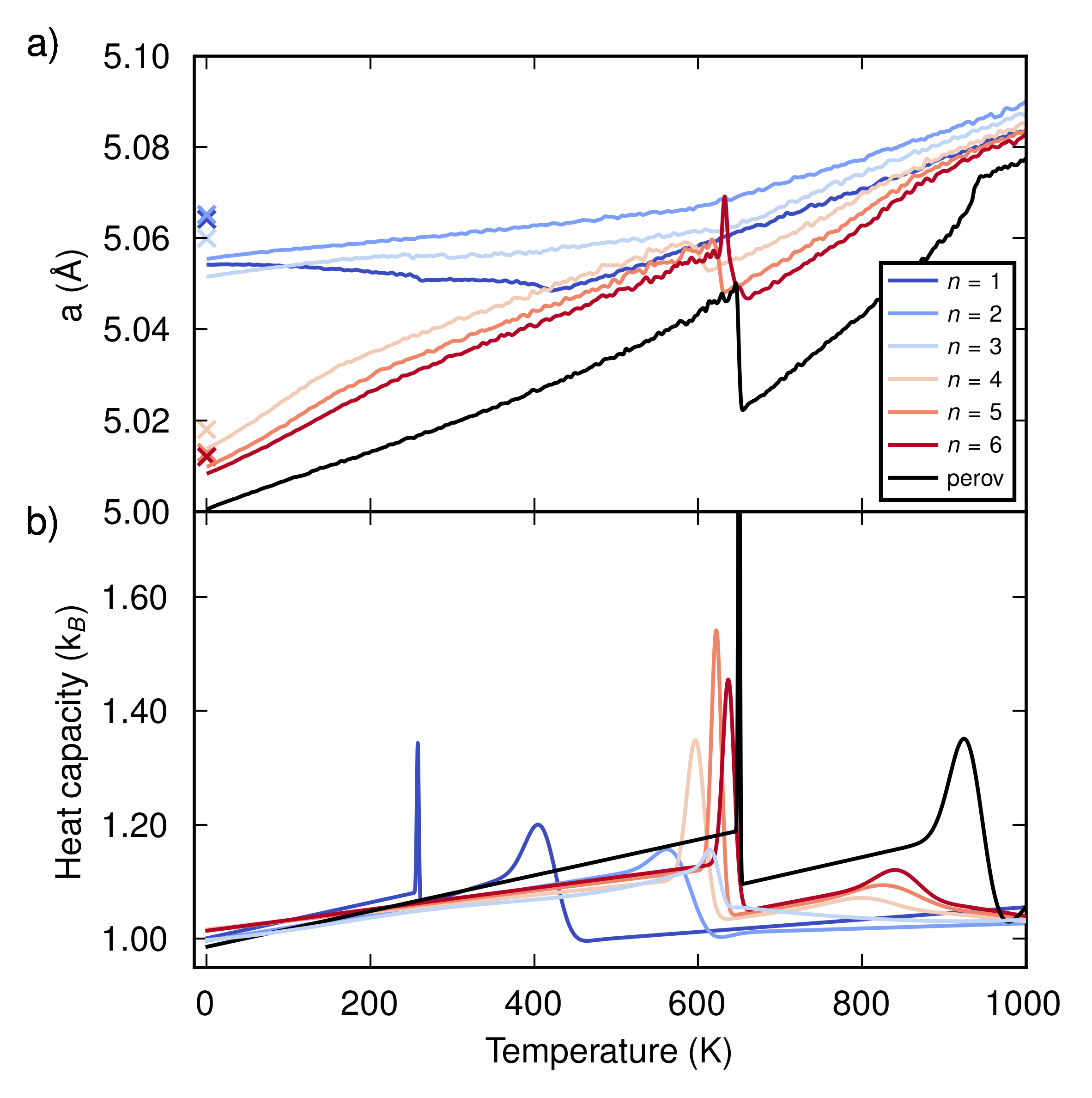}
\caption{
    Neuroevolution potential heating simulations for \ce{Ba_{n+1}Zr_nS_{3n+1}}:
    a) in-plane lattice parameters. Scatter points at \SI{0}{\kelvin}  correspond to the lattice parameters from HSE06 relaxations. 
    Note the perovskite lattice parameter is for a conventional cubic cell; b) heat capacities. A sharper peak in the heat capacity indicates a first-order (discontinuous) phase transition, whilst a broader peak indicates more second-order (continuous) character.  
}
\label{fig:heating}
\end{figure}

We investigated the temperature evolution of \gls{rp} structures via \gls{md} heating simulations starting from the ground state polymorphs at \qty{0}{\kelvin} (\autoref{fig:heating}). 
Lattice parameters cluster near \qty{5.06}{\angstrom} for low-$n$ phases and around \qty{5.01}{\angstrom} for high-$n$, with the latter contracted due to the increased in-plane tilting in their ground state structures.
Inspection of heat capacity peak positions and widths, along with octahedral tilt angles (Figs. S18-S23), indicates that high-$n$ compounds experience two transitions: a low-temperature, first-order transition, followed by a high-temperature, second-order transition.
As $n$ increases, there is evolution from layered to bulk-like behaviour, with transition temperatures progressively converging to the \ce{BaZrS3} perovskite limits at \qty{650}{\kelvin} and \qty{880}{\kelvin}.

\ce{Ba2ZrS4} exhibits negative in-plane thermal expansion up to the second order phase transition at \SI{420}{\kelvin} (\autoref{fig:heating}a), at which point the linear thermal expansion coefficient, $\alpha$, is \SI{-3e-6}{\per\kelvin}.
In contrast, \ce{Ba3Zr2S7} and \ce{Ba4Zr3S10} display positive in-plane expansion with $\alpha \approx \SI{3e-6}{\per\kelvin}$ at \SI{420}{\kelvin}.
These expansion coefficients are an order of magnitude smaller than those of the high-$n$ materials and \ce{BaZrS3}, with $\alpha \approx \SI{2e-5}{\per\kelvin}$.
Negative thermal expansion has been previously reported for RP oxides
with $n = 1$ and $n=2$, where the sign and magnitude of thermal expansion are highly sensitive to small changes in symmetry \cite{Ablitt2017}.
Recent quasi-harmonic calculations predict a negative volumetric thermal expansion in \ce{Ba3Zr2S7} at ambient pressure and below \SI{10}{\kelvin} \cite{Koocher2025tunable}.

\begin{figure}
\centering
\includegraphics[width=\linewidth]{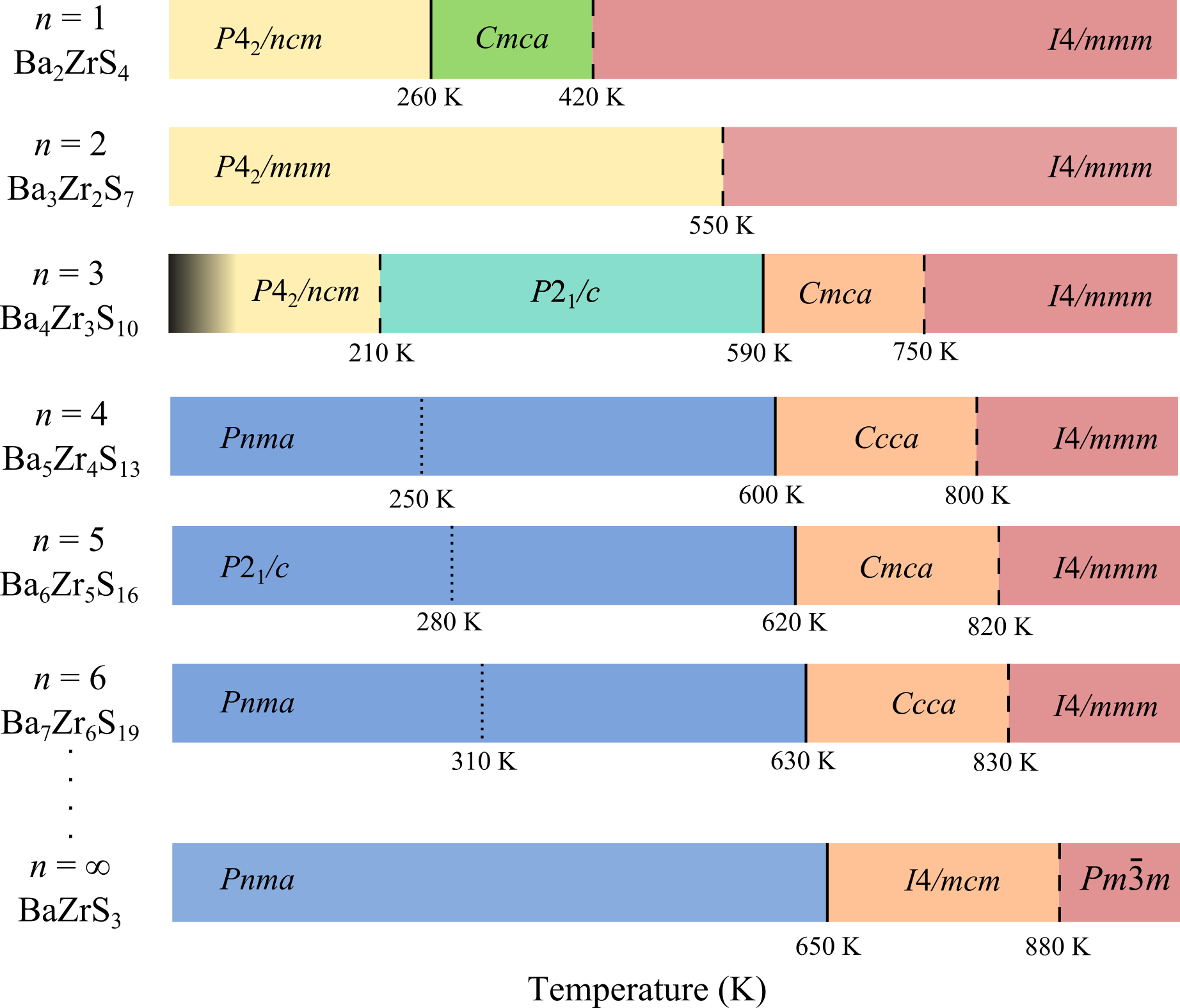}
\caption{
Summary of space groups, tilt patterns, and phase transition temperatures for \ce{Ba_{n+1}Zr_nS_{3n+1}}. Bars are colour-coded by tilt pattern; see Table S3 for Aleksandrov tilt notation. Space groups, labelled in text, depend on both the tilt pattern and the parity of $n$. Solid, dashed, and dotted lines mark first-order, second-order, and surface transitions, respectively. 
Gradient fade on $n=3$ indicates a possible low-temperature \hmn{C2/c} phase.
}
\label{fig:phase_diagram}
\end{figure}

\begin{figure*}
\centering
\includegraphics[width=\linewidth]{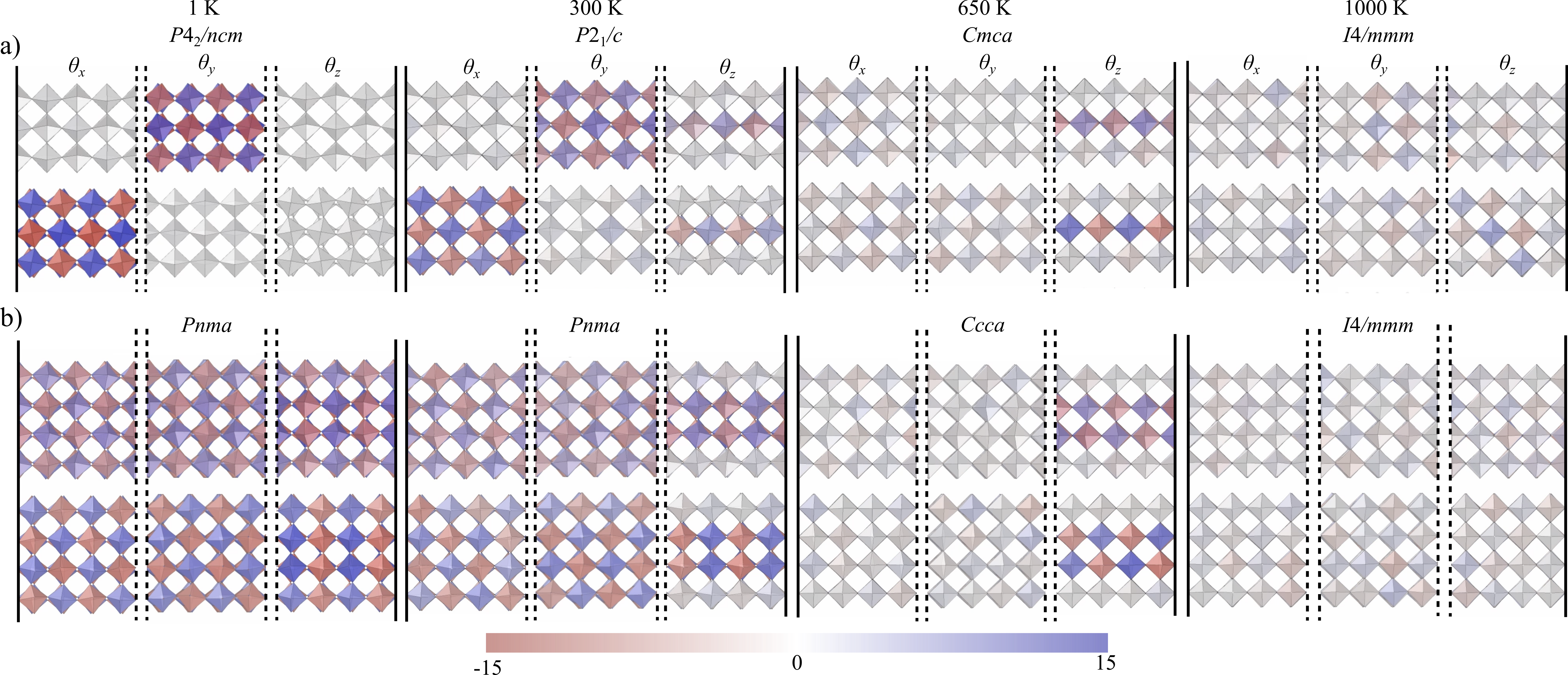}
\caption{
    Snapshots of Euler angles ($\theta_x$, $\theta_y$, $\theta_z$) for \ce{Ba_{n+1}Zr_nS_{3n+1}} for a) $n=3$ and b) $n=4$.
    Octahedra are colour-coded by the magnitude of their Euler tilt angles. 
    Solid lines separate structures by temperature, and dashed lines distinguish tilt axis.
    Structures are viewed along the $x$-axis for $\theta_x$ and $\theta_z$, and along the $y$-axis for $\theta_y$.
}
\label{fig:RP_3_RP_4_tilts}
\end{figure*}

A summary of all predicted polymorphs and transition temperatures for the \gls{rp} series and \ce{BaZrS3} is shown in \autoref{fig:phase_diagram}. 
For \gls{rp} materials with $n = 1$–3, three distinct structural evolution pathways are observed. In contrast, all phases with $n\geq4$ follow an identical sequence of tilt patterns, matching that of the parent perovskite. 
By \qty{830}{\kelvin}, all $n$-values have transitioned into the untilted \hmn{I4/mmm} parent phase, which is equivalent to the high-temperature cubic \hmn{Pm-3m} structure of \ce{BaZrS3} perovskite \cite{kayastha2025octahedral}.

Comparison against a wider range of octahedral tilt patterns catalogued by Liu et al. reveals one additional low-temperature configuration in \ce{Ba4Zr3S10}, $0\phi0~0\phi0$ (\hmn{C2/c}) \cite{liu2023understanding}. 
This is marginally stabilized by \SI{0.4}{\milli\electronvolt\per\atom} at \SI{0}{\kelvin}, with 
harmonic phonon calculations confirming its stability up to \SI{100}{\kelvin}. 
Kinetic barriers prevent its formation on cooling, and delay formation of the \hmn{P2_1/c} phase on heating. 
Therefore, all analyses and heating simulations are based on the \hmn{P4_2/ncm} ground state structure. 
The same high-temperature phases (\hmn{P2_1/c}, \hmn{Cmca}, and \hmn{I4/mmm}) are recovered on heating, regardless of the starting configuration.

We computed \gls{xrd} patterns for \ce{Ba2ZrS4} and \ce{Ba3Zr2S7} to compare with available experimental data \cite{niu2019crystal,saeki1991preparation} (Figs. S30 and S31). 
For \ce{Ba3Zr2S7}, the simulated pattern shows excellent agreement with experiment, confirming the formation of the \hmn{P4_2/mnm} phase at \qty{300}{\kelvin} and validating our computational approach.  
In \ce{Ba2ZrS4}, two phase transitions occur near room temperature. 
The \hmn{Cmca} to \hmn{I4/mmm} transition is second-order, leading to a gradual reduction of the soft mode \gls{xrd} peaks, and complicating phase identification. 
We therefore tentatively assign the room-temperature structure as \hmn{Cmca}. 
Additional single-crystal \gls{xrd} measurements, complemented by Raman spectroscopy, would help confirm this assignment more definitively.

Both the \hmn{P4_2/ncm} to \hmn{Cmca} transition in \ce{Ba2ZrS4} and the \hmn{P4_2/ncm} to \hmn{P2_1/c} transition in \ce{Ba4Zr3S10} correspond to a reduction in symmetry with increasing temperature. 
Such ascending symmetry breaking in a displacive phase transition is unusual, as higher-symmetry phases typically stabilize at elevated temperatures \cite{Orobengoa2012}, and generally indicate the presence of competing interactions with similar energy scales near the transition boundary \cite{Christensen2018emergent}. 
In \ce{Ba2ZrS4}, the symmetry lowering occurs via a first-order transition driven by enhanced in-plane octahedral tilting (Fig. S24a), coinciding with negative in-plane thermal expansion, whereas in \ce{Ba4Zr3S10} it notably occurs via a continuous transition. 

To probe the microscopic mechanism in \ce{Ba4Zr3S10}, we analyse the layer-resolved tilt angles of the \ce{ZrS6} octahedra along each axis at four different temperatures (\autoref{fig:RP_3_RP_4_tilts}a). 
At \qty{1}{\kelvin}, the space group \hmn{P4_2/ncm} with a tilt pattern $\phi00~0\Bar{\phi}0$. By \qty{210}{\kelvin}, the system has transitioned to a lower-symmetry phase,  \hmn{P2_1/c}, through the introduction of z-axis tilting confined to the central layer of each slab.
Similar interface-dependent tilt effects, where the tilt angle varies with distance from the rocksalt layer, have been reported in simulations of hybrid halide \gls{rp} materials \cite{fransson2024impact}; to our knowledge, this is the first observation of such behaviour in an inorganic \gls{rp} system.

We also analyse tilting in \ce{Ba5Zr4S13} as an example of large-$n$ behaviour (\autoref{fig:RP_3_RP_4_tilts}b). 
At \qty{1}{\kelvin}, the tilt configuration is $\phi\phi\psi_z ~\phi\phi\Bar{\psi}_z$ (\hmn{Pnma}). 
By \qty{250}{\kelvin}, tilting along the $z$-axis at the perovskite-rocksalt interface is suppressed, although the space group remains unchanged (we refer to this as a ``surface transition'').
This analysis demonstrates that the perovskite-rocksalt interface impacts on RP structural behaviour in a number of ways.
For high-$n$ \gls{rp} materials it enables the surface transitions and layer-dependent tilt patterns, whilst for low-$n$  materials, it hinders octahedral tilting at low temperature, so that tilts within a slab are along a single axis only.

To better understand these interface effects, we examine cation displacements and their coupling to octahedral tilts. 
Out-of-plane Ba displacements in the interior layers, as well as Zr displacements from their \hmn{I4/mmm} high-symmetry positions, are strongly suppressed (Fig. S25). 
In-plane Ba displacements are coupled to octahedral tilts and preserve the symmetry of the space groups (Fig. S26–S29).

\begin{figure*}
\centering
\includegraphics[width=\linewidth]{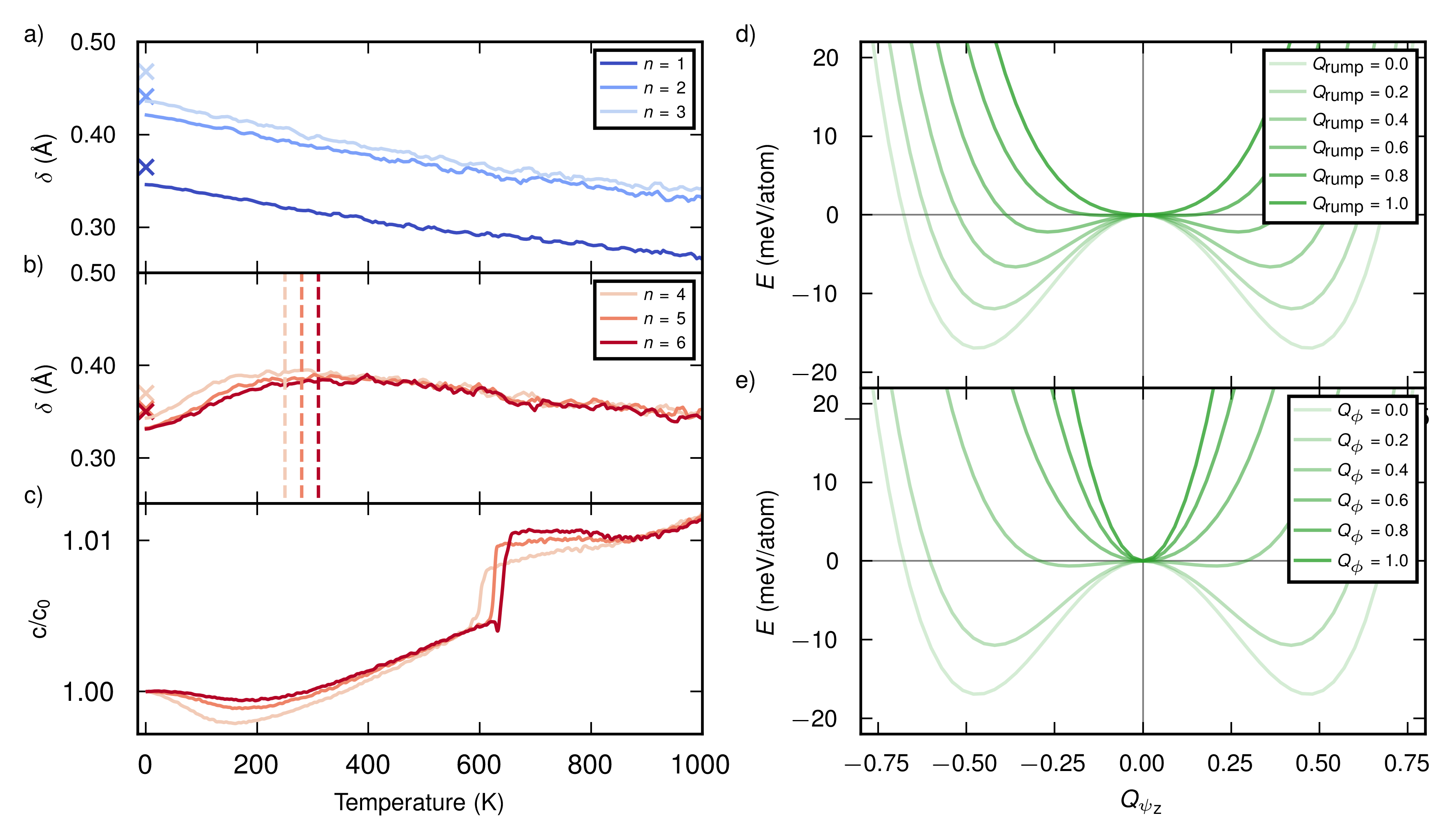}
\caption{
    Rumpling amplitudes ($\delta$) in Ba$_{n+1}$Zr$_n$S$_{3n+1}$ \gls{rp} materials with a) $n=1$ to 3 and b) $n = 4$–6. Scatter points at \SI{0}{\kelvin} correspond to the rumpling amplitudes from HSE06 relaxations. Surface transition temperatures are indicated by dashed vertical lines. 
    c) Normalised out-of-plane lattice parameters for $n = 4$–6. 
    One-dimensional potential energy surfaces for out-of-plane, in-phase octahedral tilts ($Q_{\psi_z}$) in \ce{Ba4Zr3S10} ($n=3$) are shown as a function of d) rumpling and e) in-plane, out-of-phase tilt amplitudes ($Q_\phi$). Distortions are illustrated in \autoref{fig:crystal_structures_phonons}.
}
\label{fig:rumpling}
\end{figure*}

At the perovskite-rocksalt interface we observe out-of-plane Ba displacements, so that the Ba and S ions no longer lie in the same plane (\autoref{fig:crystal_structures_phonons}d). 
This distortion is termed ``rumpling'' in the literature. 
We predict ground state rumpling amplitudes consistent with other materials that have uncharged AX rocksalt layers (\autoref{fig:rumpling}a and \autoref{fig:rumpling}b) \cite{Zhang2020unraveling}. 
We also find competition between out-of-plane octahedral rotations and rumpling, as has been previously reported for \gls{rp} materials \cite{birol2011interface, stone2016atomic, Zhang2020unraveling}. 
\autoref{fig:rumpling}b shows that, for large-$n$ RP materials only, the rumpling amplitude increases as the out-of-plane tilts at the perovskite-rocksalt interface decrease (Fig. S25-S27), reaching a maximum at the surface transition where out-of-plane tilts vanish. 
The increase in rumpling amplitude is also accompanied by negative out-of-plane thermal expansion, as shown in \autoref{fig:rumpling}c. 

The inverse coupling between octahedral rotations and rumpling is further supported by analysis of the corresponding phonon modes: large rumpling amplitudes stabilise structures with zero out-of-plane tilting ($Q_{\psi_z}=0$), while decreasing rumpling amplitudes allows the tilt amplitude to grow (\autoref{fig:rumpling}d). 
The competition arises because both distortions adjust Ba–S distances; out-of-plane tilting shortens Ba–S bonds, leading to over-coordination of Ba when combined with rumpling (see Table S4). Rumpling and its competition with octahedral tilting thus provides an understanding of why surface transitions occur in these materials.

The unusual symmetry-lowering \hmn{P4_2/ncm} to \hmn{P2_1/c} phase transition in \ce{Ba4Zr3S10} also arises from the delicate interplay between in-plane and out-of-plane octahedral tilts coupled with rumpling. 
In the ground state, large in-plane tilt amplitudes $Q_{\phi}$ suppress the out-of-plane tilt $\psi_z$ (\autoref{fig:rumpling}e). 
As temperature increases, $Q_{\phi}$ decreases, making the $\psi_z$ tilt energetically favourable and lowering the overall symmetry. 
The outermost perovskite layers remain untilted due to rumpling at the perovskite–rocksalt interface. 
In \gls{rp} materials with $n<3$, all perovskite layers are adjacent to rocksalt layers, suppressing $\psi_z$ tilts and associated symmetry-lowering transitions. 
For $n>3$, $\psi_z$ tilts are already active at \SI{0}{\kelvin}, preventing symmetry reduction. 
Thus the $n=3$ material represents a special intermediate case, featuring suppressed $\psi_z$ tilts at low temperature alongside an internal perovskite layer.

In this work, we have developed a high-accuracy machine-learned interatomic potential to systematically explore the temperature- and $n$-dependent structural evolution of the \gls{rp} chalcogenide series Ba$_{n+1}$Zr$_n$S$_{3n+1}$.
Our simulations reveal a rich and diverse set of polymorphs exhibiting notable phenomena: negative thermal expansion, symmetry-lowering phase transitions, and complex, layer-dependent tilt patterns. 
Analysis shows that this structural diversity arises from the interplay between octahedral tilt modes and rumpling distortions at the perovskite-rocksalt interface. 
The methodology can be extended to investigate phase stabilization strategies, such as selective A-site doping at the perovskite–rocksalt interface, or applied to understand the dynamics of related materials with displacive structural phase transitions.

\section*{Acknowledgments}
P.K. gratefully acknowledges funding through the Turing Scheme, which enabled a research visit to Chalmers University of Technology, and support from the UK Engineering and Physical Sciences Research Council (EPSRC) CDT in Renewable Energy Northeast Universities (ReNU) via Grant EP/S023836/1. This work also benefits from COST Action RenewPV CA21148, supported by COST (European Cooperation in Science and Technology), as well as funding from the Swedish Research Council (Nos. 2020-04935 and 2021-05072) and the Knut and Alice Wallenberg Foundation (No.~2024.0042).

Computational resources were provided by the Oswald High-Performance Computing Facility at Northumbria University, the ARCHER2 UK National Supercomputing Service through the HEC Materials Chemistry Consortium (EPSRC EP/X035859), and the UK Materials and Molecular Modelling Hub (EPSRC EP/T022213, EP/W032260, EP/P020194). Additional support came from the National Academic Infrastructure for Supercomputing in Sweden (NAISS) at NSC, PDC, and C3SE, partially funded via the Swedish Research Council (Grant 2022-06725), and the Berzelius supercomputing resource, provided by the Knut and Alice Wallenberg Foundation at NSC.

We thank Kevin Ye and Shanyuan Niu for providing experimental X-ray diffraction data, and Kevin Ye, Shanyuan Niu, Emma McCabe, and Nick Bristowe for valuable discussions related to this work.

\section*{Data Availability Statement}
The \gls{nep} models produced in this study are publicly accessible via Zenodo: \url{https://doi.org/10.5281/zenodo.15829577}. The \gls{dft} output data have been deposited in the NOMAD repository: \url{https://doi.org/10.17172/NOMAD/2025.07.11-1}. A paper-specific repository, containing Python code to reproduce the figures and analyses, is available at \url{https://github.com/NU-CEM/2025_RP_phases}.
\bibliographystyle{apsrev4-2}
\bibliography{prl_reformatted}

\end{document}


\maketitle

\tableofcontents{}

\newpage
\section{Methods}

\subsection{Neuroevolution potential method}
An iterative strategy was employed to construct the neuroevolution potential outlined in Ref.~\citenum{fransson2023phase}.
The \textsc{gpumd} package\cite{FanZenZha21, Fan22, FanWanYin22}  (version 3.9.5) was used to build the \gls{nep} model and run the \gls{md} simulations.
The \ase{} \cite{Larsen2017} and \calorine{} \cite{calorine} packages were used to prepare the training structures, set up \gls{md} simulations and post-process the results. Random displacements were generated using the \hiphive{} package \cite{EriFraErh19}.

The initial training set contains strained primitive structures and rattled supercells of the perovskite and the $n=1 ~\text{to} ~6$ RP phases. 
The Ruddlesden-Popper training structures include ideal, distorted, and deformed unit cells and supercells for $n=1$ to $n=6$, along with the 33 tilt pattern structures defined by Aleksandrov for $n=1$ to $n=10$. Molecular dynamics (MD) structures were included for temperatures up to \qty{1000}{\kelvin}.
The different phases considered for the perovskite are described in Ref.~\citenum{kayastha2025octahedral}. 
The final training set consists of 1375 structures.

The initial model was trained using \gls{dft} data generated with the PBEsol exchange-correlation functional \cite{perdew2008restoring}. 
For each structure in the training set, formation energy (relative to the elemental phases), atomic forces and stress tensors were evaluated using \gls{dft}. 
This model was then used to run \gls{md} simulations in the NPT ensemble, over a temperature range of \qtyrange{1}{1000}{\kelvin} with varying supercell sizes, containing between 20 and 500 atoms.
The model was retrained after adding snapshots from randomly selected structures from the \gls{md} runs. 

On the final training set, we used higher accuracy HSE06 functional\cite{krukau2006influence} to calculate energies, forces, and stress tensors. 
Stress tensors for structures containing more than 60 atoms were not evaluated as the memory requirements for these calculations were prohibitively large.

\subsection{Density functional theory calculations}
Formation energies, forces, and stress tensors used in the training set were calculated using the all-electron numeric atom-centered orbital code FHI-aims \cite{blum2009ab}.
Due to large memory requirements, the stress tensors are not calculated for structures larger than 60 atoms in the unit cell. 
FHI-vibes \cite{knoop2020fhi} was used for pre and post-processing of \gls{dft} data.  All \gls{dft} calculations used the $\textit{light}$ basis set and a Monkhorst-Pack $k$-point mesh with a minimum $k$-spacing of \qty{0.2}{\per\angstrom}. For single-point calculations, the charge density was converged to an accuracy of \num{e-6}, forces to \qty{e-5}{\electronvolt\per\angstrom}, and stresses to \qty{e-4}{\electronvolt\per\angstrom\cubed}. 
Geometry relaxations were carried out using the symmetry-constrained relaxation scheme as implemented in \ase{} \cite{Larsen2017}, until the maximal force component was below \qty{e-3}{\electronvolt\per\angstrom}.
Harmonic phonon dispersions at \qty{0}{\kelvin} using the \textsc{phonopy} package \cite{togo2023first} with a \numproduct{2x2x2} supercell and a \qty{0.01}{\angstrom} displacement distance. Forces for these phonon calculations were converged below \qty{e-3}{\electronvolt\per\angstrom}. 

\subsection{Molecular dynamics simulations}
To perform the heating runs, the \gls{md} simulations were carried out using the \textsc{gpumd} software \cite{FanWeiVie2017} with a timestep of \qty{1}{\femto\second}.
The \ase{} \cite{Larsen2017} and calorine \cite{calorine} packages were used to set up the \gls{md} simulations and post-process the results.
Heating and cooling simulations were run in the NPT ensemble between \qtyrange{0}{1200}{\kelvin} for \qty{50}{\nano\second} using a supercell consisting of about \num{40000} atoms.
The potential energy and lattice parameters were recorded every \qty{100}{\femto\second} to discern phase transitions.

Additional cooling simulations were performed for both the Ruddlesden-Popper series and the perovskite phase, starting from the ideal \hmn{I4/mmm} and \hmn{Pm\overline{3}m} structures, respectively. For RP phases with $n \geqslant 6$ and for the perovskite, we observe significant hysteresis during cooling, where the system becomes kinetically trapped in a metastable phase and does not undergo the expected first-order transition, persisting in this state down to \qty{0}{\kelvin}.

The heat capacities across the different phase transitions were calculated by fitting the potential energy and the temperature data from \gls{md} to an analytical expression and then taking its derivative as done in Ref.~\citenum{fransson2024impact}.
The analytical function for the energy is a polynomial in temperature for each phase, splined together with an S-like error-function, where the location and width of this splining are also fit parameters.
This ensures that the analytical form can capture sharp discontinuous changes in the energy as well as broad continuous changes in the energy across phase transitions, and therefore allows a more reliable estimate of the heat capacity to be obtained compared to direct numerical differentiation.

To identify the finite-temperature polymorphs stabilized during heating, atomic displacements were projected onto the in-phase and out-of-phase octahedral tilt eigenmodes of the ideal \hmn{I4/mmm} phase (Figs. S6-S17). These mode projections were evaluated separately for the two inequivalent slabs within the Ruddlesden-Popper structure. For the perovskite phase, projections were instead based on the phonon eigenvectors of the \hmn{Pm\overline{3}m} structure, following the methodology in Refs.\citenum{FraRosEriRahTadErh23}. The space groups of RP structures exhibiting suppressed surface-layer tilting were assigned by constructing idealized structural models with the same octahedral tilt as observed in \gls{md}, and evaluating them with \textsc{spglib} \cite{Togo2024}.

Structural visualizations and analysis were performed using the \ovito ~package \cite{stukowski2009visualization}.

\newpage
\section{Model validation}
\label{snote:validation_of_NEP}

The final NEP model achieved root mean squared errors of \qty{1.8}{\milli\electronvolt\per\atom} for formation energies, \qty{68.4}{\milli\electronvolt\per\angstrom} for atomic forces, and \qty{26.1}{\milli\electronvolt\per\angstrom^3\per\atom} for virial stresses. Loss curves, root mean square errors and parity plots are provided below.

\begin{figure}[H]
    \centering
    \includegraphics[width=0.7\linewidth]{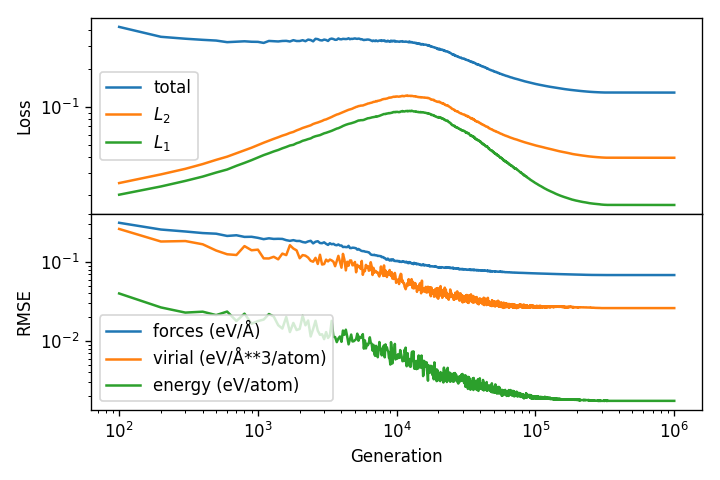}
    \caption{Loss curves and root mean square errors (RMSE) of energies, forces, and virials for the full model.}
    \label{fig:enter-label}
\end{figure}
\begin{figure}[H]
    \centering
    \includegraphics[width=\linewidth]{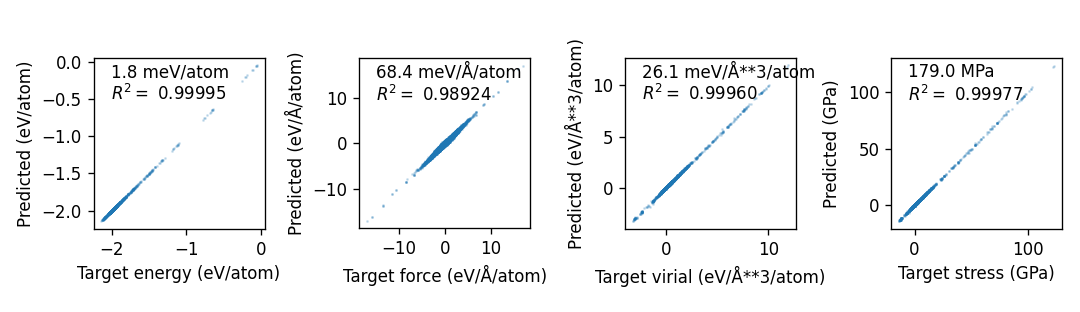}
    \caption{Parity plots for the full model. Structures for which the stress tensors have not been evaluated with DFT are not included in the virial and stress tensor plots. }
    \label{fig:enter-label}
\end{figure}

The \qty{0}{\kelvin} harmonic phonon spectra for Ruddlesden-Popper materials in the high-temperature phase are displayed below. The solid blue lines correspond to phonons generated by evaluating forces from the NEP model, and the dashed black lines are phonons generated by evaluating forces from DFT. The largest discrepancies correspond to high-frequency modes associated with the sulfur species.\cite{wu2023ultralow} 
\begin{figure}[H]
    \centering
    \includegraphics[width=0.45\linewidth]{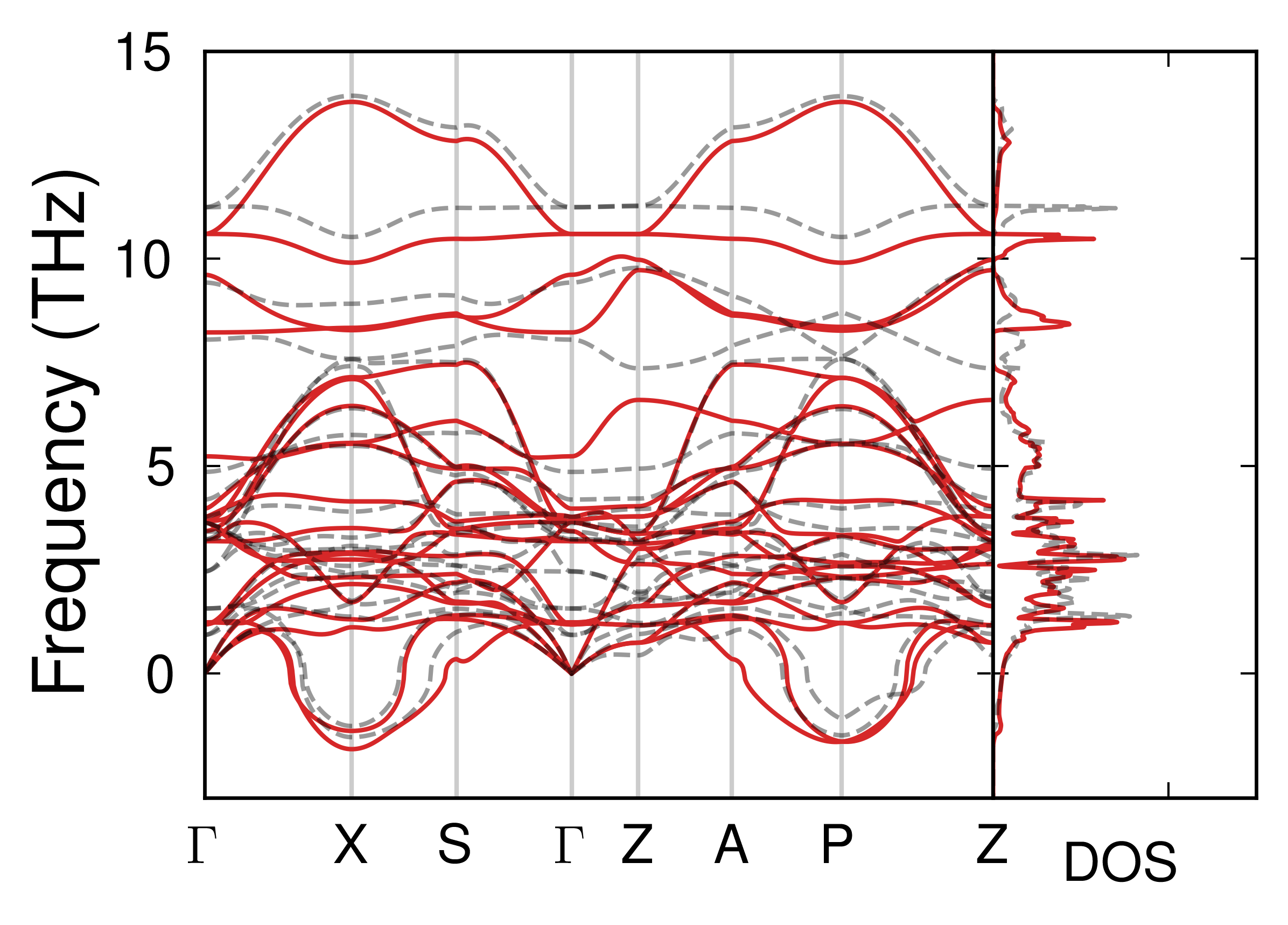}
    \caption{Phonons and DOS of the $n=1$ RP in the \hmn{I4/mmm} phase using NEP (solid red) and DFT (dashed black)}
    \label{fig:phonons_n_1}
\end{figure}
\begin{figure}[H]
    \centering
    \includegraphics[width=0.45\linewidth]{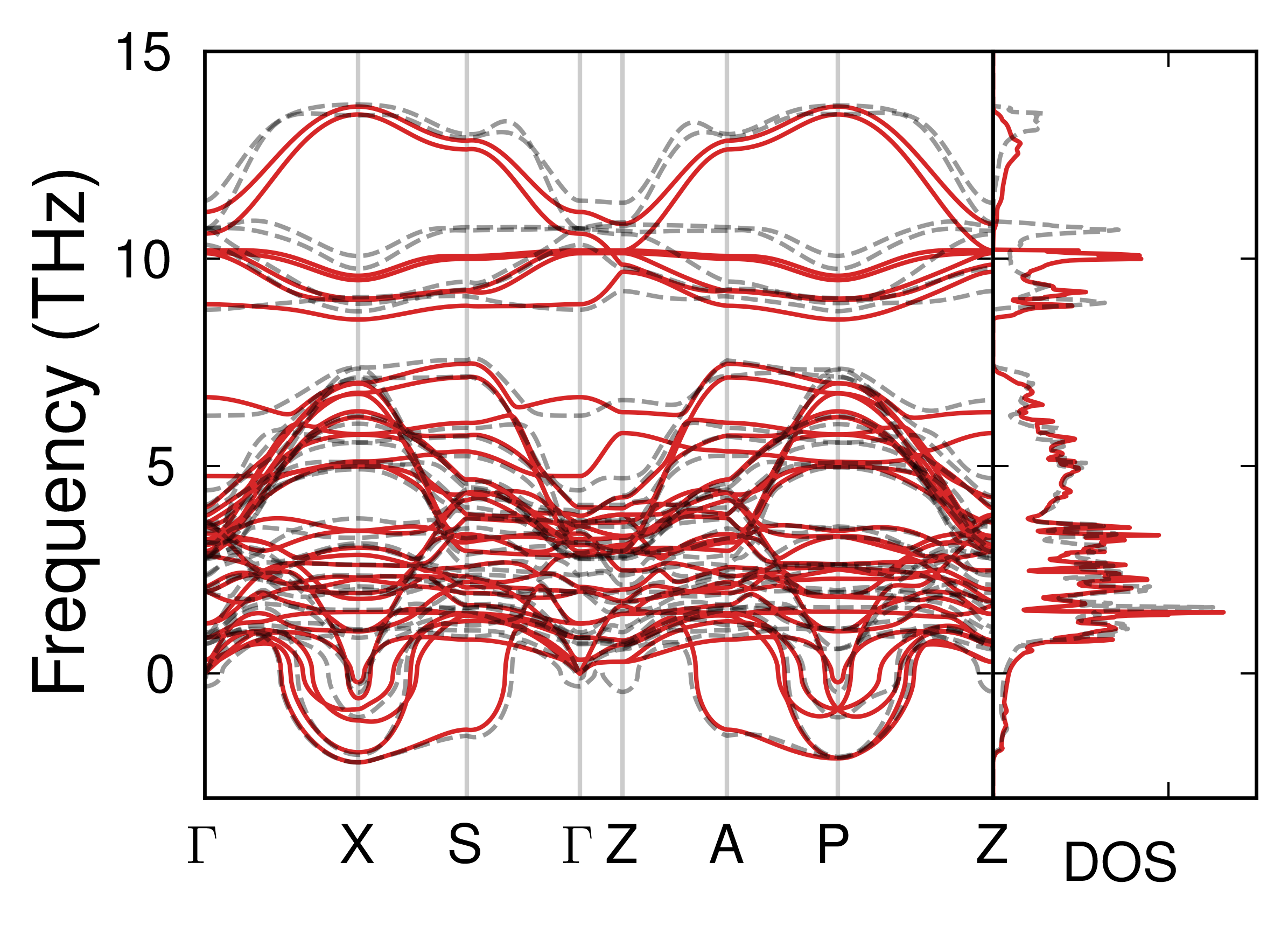}
    \caption{Phonons and DOS of the $n=2$ RP in the \hmn{I4/mmm} phase using NEP (solid red) and DFT (dashed black)}
    \label{fig:phonons_n_2}
\end{figure}
\begin{figure}[H]
    \centering
    \includegraphics[width=0.45\linewidth]{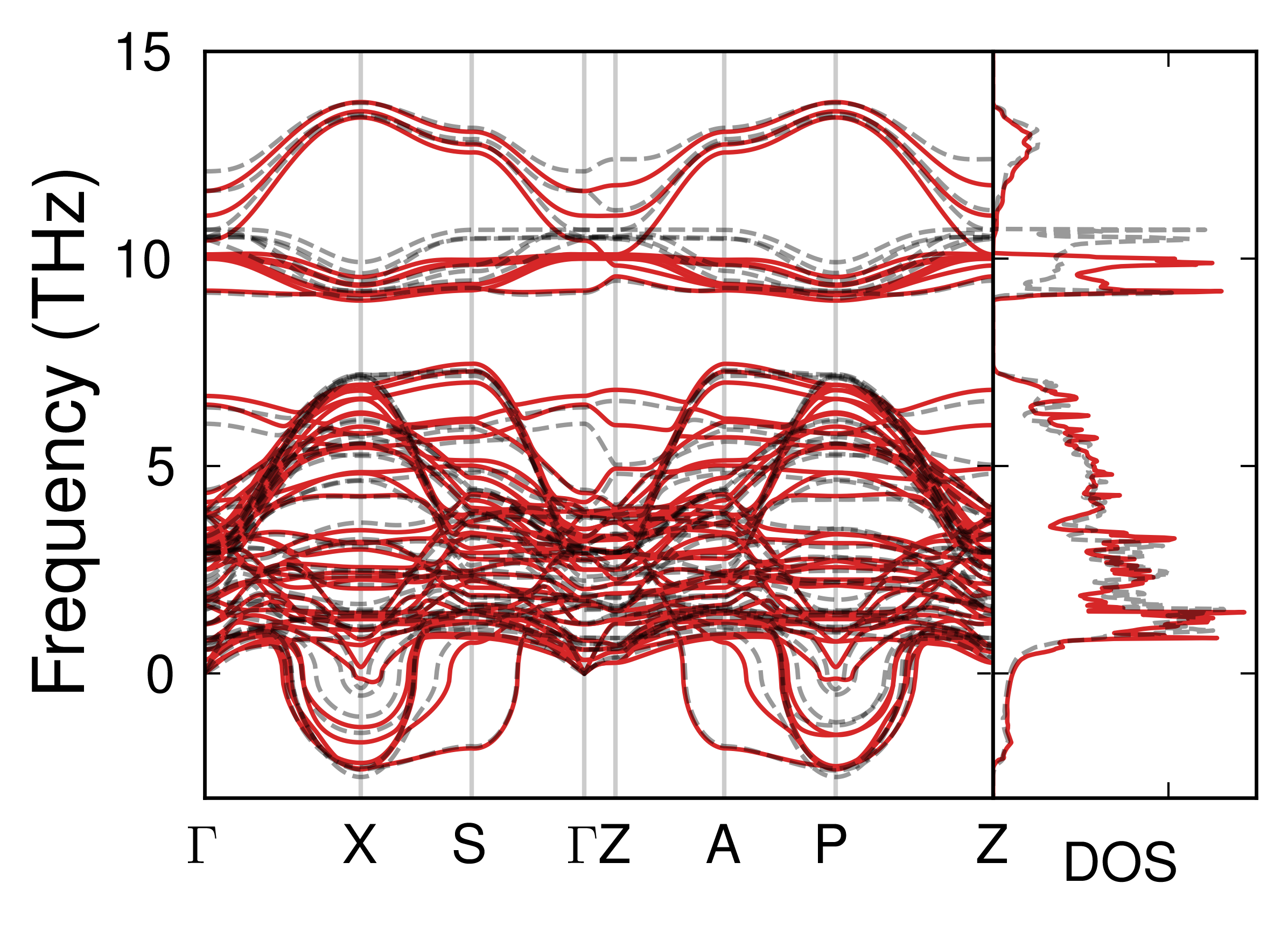}
    \caption{Phonons and DOS of the $n=3$ RP in the \hmn{I4/mmm} phase using NEP (solid red) and DFT (dashed black)}
    \label{fig:phonons_n_3}
\end{figure}

Key sources of error in the predicted phase transition temperatures of halide perovskites, using the same NEP-based methodology, have been analysed in a previous publication \cite{fransson2023phase}. This includes analysing the impact of heating and cooling rates, supercell size, model uncertainty and exchange-correlation functional. The error associated with simulation parameters can be minimised through careful convergence, as we have done here, however the largest source of error in most studies is likely the exchange-correlation functional, with phase transition temperatures varying by as much as \SI{150}{\kelvin} (less for closely related functionals describing the same physics). In this work we train our NEP on a more accurate electronic structure method (the HSE06 functional \cite{heyd2003hybrid}), thus reducing the largest source of error. Moving to a higher level of theory (e.g. RPA) may show another shift in transition temperature, however it is currently intractable to evaluate the required number of structures and system sizes with a method beyond hybrid DFT. We note that the difference between the predicted and experimentally measured orthorhombic-to-tetragonal phase transition temperature in \ce{BaZrS3}, using the same HSE06-NEP methodology as outlined here, is only \SI{50}{\kelvin} \cite{kayastha2025octahedral}, which provides an estimate of the expected error bar in this work. 

\newpage

\section{Ground state energies}
NEP-calculated energies of the 33 unique octahedral tilt patterns, after geometry relaxation with constrained symmetry. Aleksandrov notation is used to denote the out-of-phase ($\phi$) and in-phase ($\psi$) octahedral tilts along each axis, with an overbar to indicate that the tilt patterns between adjacent slabs are in opposite directions \cite{aleksandrov1987successive}. Subscripts are used to differentiate between tilts of different magnitude, where e.g. $\phi_1 \neq \phi_2$. The energies are specified relative to the ground state (which is indicated in bold). Some tilt patterns have the same energy. In this case, the lower-symmetry phase is unstable and has relaxed to the higher symmetry phase. The ground state structures have been confirmed through DFT electronic relaxations with the HSE06 functional.

\begin{table}[H]
    \centering
    \begin{tabular}{llccc}
 space group & Slab I Slab II & $n = 1$ & $n = 3$& $n=5$\\
 \hline
\hmn{I4/mmm}     &  000                             000                                                              &  0.00719            &  0.01314            &  0.01909            \\
\hmn{Cmca}       &  00$\psi_z$                    00$\psi_z$                                                     &  0.00719            &  0.00516            &  0.00737            \\
\hmn{P4/mbm}     &  00$\psi_z$                    000                                                              &  0.00719            &  0.01030            &  0.01451            \\
\hmn{Pbam}       &  00$\psi$$_{z_1}$                00$\psi$$_{z_2}$                                                 &  0.00719            &  0.00516            &  0.00737            \\
\hmn{Cmca}       &  00$\phi_z$                    00$\phi_z$                                                     &  0.00719            &  0.00516            &  0.00467            \\
\hmn{P4/mbm}     &  00$\phi_z$                    000                                                              &  0.00719            &  0.01030            &  0.01328            \\
\hmn{Pbam}       &  00$\phi$$_{z_1}$                00$\phi$$_{z_2}$                                                 &  0.00719            &  0.00516            &  0.00467            \\
\hmn{P4_2/nnm}   &  $\phi$00                    0$\phi$0                                                     &  0.00456            &  0.00245            &  0.00441            \\
\hmn{Cccm}       &  $\phi$$\phi$0           $\phi$$\phi$0                                            &  0.00486            &  0.00416            &  0.00538            \\
\hmn{Pnnn}       &  $\phi$$_{1}$$\phi$$_{2}$0   $\phi$$_{2}$$\phi$$_{1}$0                                    &  0.00456            &  0.00245            &  0.00437            \\
\hmn{P4_2/ncm}   &  $\phi$00                    0$\Bar{\phi}$0                                          &  \textbf{0.00000}   &  \textbf{0.00000}   &  0.00287            \\
\hmn{Cmca}       &  $\phi$$\phi$0           $\Bar{\phi}$$\Bar{\phi}$0                      &  0.00082            &  0.00194            &  0.00399            \\
\hmn{Pccn}       &  $\phi$$_{1}$$\phi$$_{2}$0   $\Bar{\phi}$$_{2}$$\phi$$_{1}$0                         &  0.00000   & 0.00000   &  0.00284            \\
\hmn{P2_1/c}     &  $\phi$0$\psi_z$           0$\phi$$\psi_z$                                            &  0.00000   &  0.00000   &  0.00005            \\
\hmn{P2_1/c}     &  $\phi$0$\phi_z$           0$\phi$$\phi_z$                                            &  0.00000   &  0.00000   &  0.00286            \\
\hmn{P2/c}       &  $\phi$0$\psi_z$           0$\Bar{\phi}$$\psi_z$                                 &  0.00456            &  0.00245            &  0.00114            \\
\hmn{P2/c}       &  $\phi$0$\phi_z$           0$\Bar{\phi}$$\phi_z$                                 &  0.00456            &  0.00245            &  0.00492            \\
\hmn{C2/m}       &  $\phi$0$\psi_z$           0$\phi$0                                                     &  0.00000   &  0.00000   &  0.00287            \\
\hmn{C2/m}       &  $\phi$0$\phi_z$           0$\phi$0                                                     &  0.00000   &  0.00000   &  0.00287            \\
\hmn{C2/m}       &  $\phi$0$\psi_z$           0$\Bar{\phi}$0                                          &  0.00000   &  0.00245            &  0.00439            \\
\hmn{C2/m}       &  $\phi$0$\phi_z$           0$\Bar{\phi}$0                                          &  0.00000   &  0.00245            &  0.00439            \\
\hmn{C2/c}       &  $\phi$$\phi$$\psi_z$  $\phi$$\phi$$\psi_z$                                   &  0.00486            &  0.00191            &  0.00114            \\
\hmn{C2/c}       &  $\phi$$\phi$$\phi_z$  $\phi$$\phi$$\phi_z$                                   &  0.00486            &  0.00191            &  0.00466            \\
\hmn{Pccn}       &  $\phi$$\phi$$\psi_z$  $\phi$$\phi$$\Bar{\psi}_z$                       &  0.00486            &  0.00192            &  0.00114            \\
\hmn{Pccn}       &  $\phi$$\phi$$\phi_z$  $\phi$$\phi$$\Bar{\phi}_z$                       &  0.00486            &  0.00192            &  0.00467            \\
\hmn{Pbca}       &  $\phi$$\phi$$\psi_z$  $\Bar{\phi}$$\Bar{\phi}$$\psi_z$             &  0.00082            &  0.00013            &  0.00005            \\
\hmn{Pbca}       &  $\phi$$\phi$$\phi_z$  $\Bar{\phi}$$\Bar{\phi}$$\phi_z$             &  0.00082            &  0.00013            &  0.00371            \\
\hmn{P2_1/c}     &  $\phi$$\phi$$\psi_z$  $\Bar{\phi}$$\Bar{\phi}$$\Bar{\psi}_z$ &  0.00082            &  0.00005            &  \textbf{0.00000}   \\
\hmn{P2_1/c}     &  $\phi$$\phi$$\phi_z$  $\Bar{\phi}$$\Bar{\phi}$$\Bar{\phi}_z$ &  0.00082            &  0.00005            &  0.00367            \\
\hmn{P2_1/c}     &  $\phi$$\phi$$\psi_z$  $\phi$$\phi$0                                            &  0.00486            &  0.00192            &  0.00114            \\
\hmn{P2_1/c}     &  $\phi$$\phi$$\psi_z$  $\Bar{\phi}$$\Bar{\phi}$0                      &  0.00082            &  0.00013            &  0.00005            \\
\hmn{P2_1/c}     &  $\phi$$\phi$$\phi_z$  $\phi$$\phi$0                                            &  0.00486            &  0.00191            &  0.00302            \\
\hmn{P2_1/c}     &  $\phi$$\phi$$\phi_z$  $\Bar{\phi}$$\phi$0                                 &  0.00082            &  0.00005            &  0.00195            \\

    \end{tabular}
    \caption{Space groups, tilt patterns, and NEP-calculated energies of RP structures with an odd number of layers within each slab. Energies are given in \qty{}{\electronvolt\per\atom} and referenced to the ground state.}
    \label{tab:Aleksandrov_NEP}
\end{table}
\begin{table}[H]
    \centering
    \begin{tabular}{llccc}
     space group & Slab I Slab II & $n = 2$ & $n = 4$& $n=6$\\
 \hline
  \hmn{I4/mmm}          &  000                             000                                                              &  0.01054           &  0.01661           &  0.02089           \\ 
 \hmn{Cmca}      &  00$\psi_z$                    00$\psi_z$                                                     &  0.01034           &  0.00921           &  0.00798           \\ 
 \hmn{P4/mbm}    &  00$\psi_z$                    000                                                              &  0.01054           &  0.01279           &  0.01575           \\ 
 \hmn{Pbam}      &  00$\psi$$_{z_1}$                00$\psi$$_{z_2}$                                                 &  0.01034           &  0.00921           &  0.00798           \\ 
 \hmn{Ccce}      &  00$\phi_z$                    00$\phi_z$                                                     &  0.00826           &  0.00482           &  0.00457           \\ 
 \hmn{P4/nbm}    &  00$\phi_z$                    000                                                              &  0.01032           &  0.01202           &  0.01419           \\ 
 \hmn{Pban}      &  00$\phi$$_{z_1}$                00$\phi$$_{z_2}$                                                 &  0.00826           &  0.00482           &  0.00457           \\ 
 \hmn{P4_2/mnm}  &  $\phi$00                    0$\phi$0                                                     &  \textbf{0.00000}  &  0.00166           &  0.00373           \\ 
 \hmn{Cmcm}      &  $\phi$$\phi$0           $\phi$$\phi$0                                            &  0.00200           &  0.00315           &  0.00457           \\ 
 \hmn{Pnnm}      &  $\phi$$_{1}$$\phi$$_{2}$0   $\phi$$_{2}$$\phi$$_{1}$0                                    &  0.00000  &  0.00166           &  0.00360           \\ 
 \hmn{P4_2/mcm}  &  $\phi$00                    0$\Bar{\phi}$0                                          &  0.00340           &  0.00356           &  0.00502           \\ 
 \hmn{Cmma}      &  $\phi$$\phi$0           $\Bar{\phi}$$\Bar{\phi}$0                      &  0.00496           &  0.00487           &  0.00573           \\ 
 \hmn{Pccm}      &  $\phi$$_{1}$$\phi$$_{2}$0   $\Bar{\phi}$$_{2}$$\phi$$_{1}$0                         &  0.00340           &  0.00356           &  0.00488           \\ 
 \hmn{Pma2}      &  $\phi$0$\psi_z$           0$\phi$$\psi_z$                                            &  0.00340           &  0.00141           &  0.00095           \\ 
 \hmn{P2/c}      &  $\phi$0$\phi_z$           0$\phi$$\phi_z$                                            &  0.00340           &  0.00354           &  0.00523           \\ 
 \hmn{Pmn2_1}    &  $\phi$0$\psi_z$           0$\Bar{\phi}$$\psi_z$                                 &  0.00000  &  0.00006           &  0.00004           \\ 
 \hmn{P2_1/c}    &  $\phi$0$\phi_z$           0$\Bar{\phi}$$\phi_z$                                 &  0.00000  &  0.00166           &  0.00431           \\ 
 \hmn{Amm2}      &  $\phi$0$\psi_z$           0$\phi$0                                                     &  0.00340           &  0.00356           &  0.00500           \\ 
 \hmn{C2/m}      &  $\phi$0$\phi_z$           0$\phi$0                                                     &  0.00340           &  0.00355           &  0.00520           \\ 
 \hmn{Amm2}      &  $\phi$0$\psi_z$           0$\Bar{\phi}$0                                          &  0.00000  &  0.00166           &  0.00371           \\ 
 \hmn{C2/m}      &  $\phi$0$\phi_z$           0$\Bar{\phi}$0                                          &  0.00000  &  0.00166           &  0.00408           \\ 
 \hmn{Cmc2_1}    &  $\phi$$\phi$$\psi_z$  $\phi$$\phi$$\psi_z$                                   &  0.00200           &  0.00006           &  0.00004           \\ 
 \hmn{C2/c}      &  $\phi$$\phi$$\phi_z$  $\phi$$\phi$$\phi_z$                                   &  0.00200           &  0.00248           &  0.00401           \\ 
 \hmn{Pnma}      &  $\phi$$\phi$$\psi_z$  $\phi$$\phi$$\Bar{\psi}_z$                       &  0.00200           &  \textbf{0.00000}  &  \textbf{0.00000}  \\ 
 \hmn{Pbcn}      &  $\phi$$\phi$$\phi_z$  $\phi$$\phi$$\Bar{\phi}_z$                       &  0.00200           &  0.00254           &  0.00405           \\ 
 \hmn{Pbcm}      &  $\phi$$\phi$$\psi_z$  $\Bar{\phi}$$\Bar{\phi}$$\psi_z$             &  0.00496           &  0.00141           &  0.00096           \\ 
 \hmn{Pcca}      &  $\phi$$\phi$$\phi_z$  $\Bar{\phi}$$\Bar{\phi}$$\phi_z$             &  0.00496           &  0.00394           &  0.00479           \\ 
 \hmn{Abm2}      &  $\phi$$\phi$$\psi_z$  $\Bar{\phi}$$\Bar{\phi}$$\Bar{\psi}_z$ &  0.00496           &  0.00141           &  0.00095           \\ 
 \hmn{P2/c}      &  $\phi$$\phi$$\phi_z$  $\Bar{\phi}$$\Bar{\phi}$$\Bar{\phi}_z$ &  0.00496           &  0.00392           &  0.00479           \\ 
 \hmn{Pmc2_1}    &  $\phi$$\phi$$\psi_z$  $\phi$$\phi$0                                            &  0.00200           &  0.00006           &  0.00004           \\ 
 \hmn{Pmc2_1}    &  $\phi$$\phi$$\psi_z$  $\Bar{\phi}$$\Bar{\phi}$0                      &  0.00496           &  0.00141           &  0.00095           \\ 
 \hmn{P2/c}      &  $\phi$$\phi$$\phi_z$  $\phi$$\phi$0                                            &  0.00200           &  0.00248           &  0.00310           \\ 
 \hmn{P2/c}      &  $\phi$$\phi$$\phi_z$  $\Bar{\phi}$$\phi$0                                 &  0.00496           &  0.00392           &  0.00399           \\ 
    \end{tabular}
    \caption{Space groups, tilt patterns, and NEP-calculated energies of RP structures with an even number of layers within each slab. Energies are given in \qty{}{\electronvolt\per\atom} and referenced to the ground state.}
    \label{tab:Aleksandrov_NEP}
\end{table}

\newpage
\section{Projection onto octahedral tilt modes}
To identify the finite-temperature polymorphs, molecular dynamic trajectories were projected onto the $\phi$ (out-of-phase) and $\psi$ (in-phase) octahedral tilt modes in the heating and cooling runs of the $n = 1$ to $6$ RP phases. The tilt components are resolved along the $x$, $y$, and $z$ directions for each slab, which is identified by subscript index. The sign of the projection amplitudes $Q$ are used to determine the relative tilt directions between slabs, with opposite tilt directions denoted with an overbar in Aleksandrov notation.

For $n=1,2,3$ the heating and cooling runs produce identical mode projection results, apart from sometimes an arbitrary flip of the sign of the mode coordinates.
However, for $n=4$ and above in the cooling runs sometimes one does not achieve perfect long-range ordering between the slabs and thus the global mode coordinate (averaged over all slabs) can take on different values or sometimes be zero while the structure still is tilted and looks similar to the structure of the heating run.

\begin{figure}[H]
    \centering
    \includegraphics[width=0.7\linewidth]{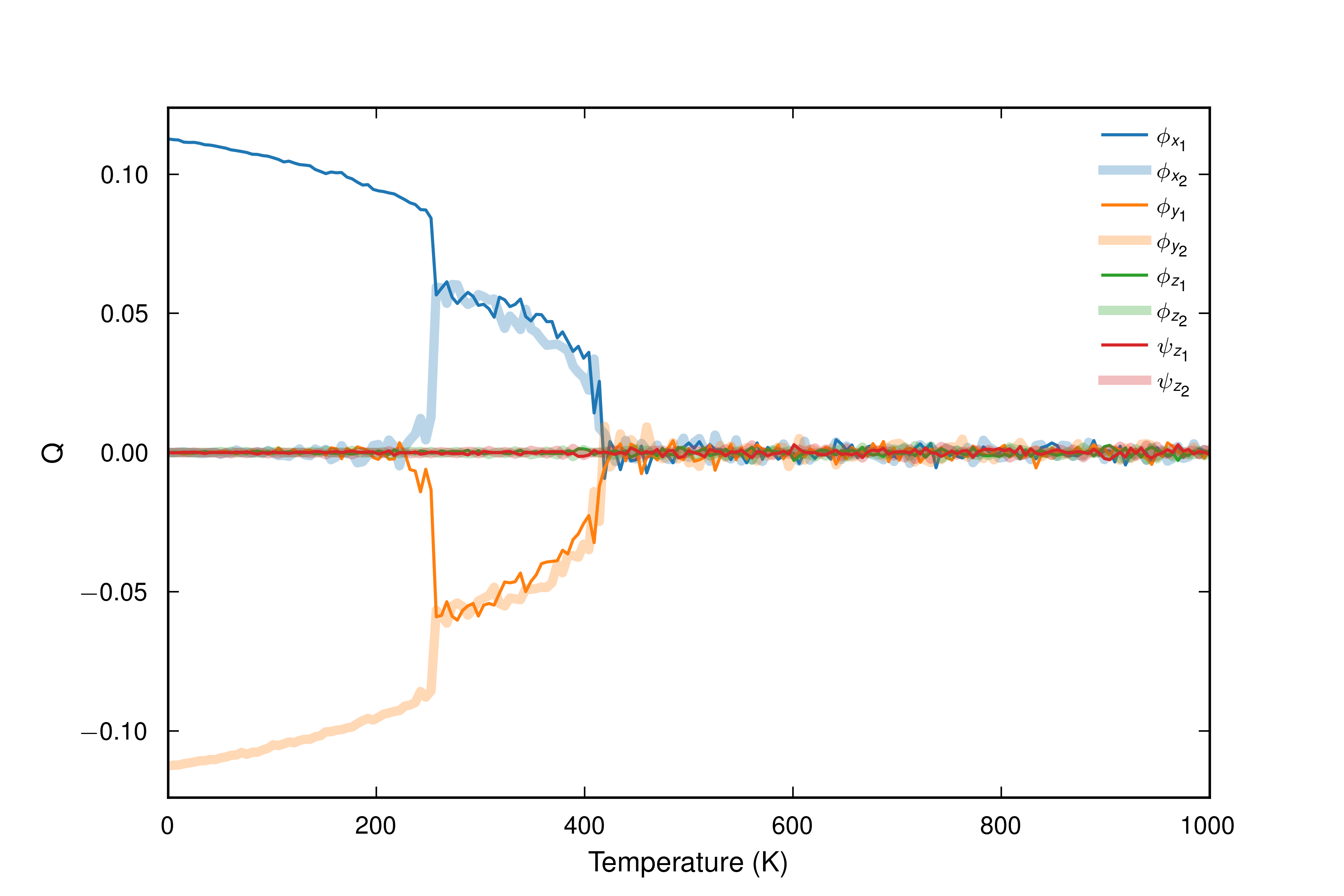}
    \caption{Mode projections in the heating run for the $n = 1$ RP phase.}
    \label{fig:n1_heating_tilt}
\end{figure}
\begin{figure}[H]
    \centering
    \includegraphics[width=0.7\linewidth]{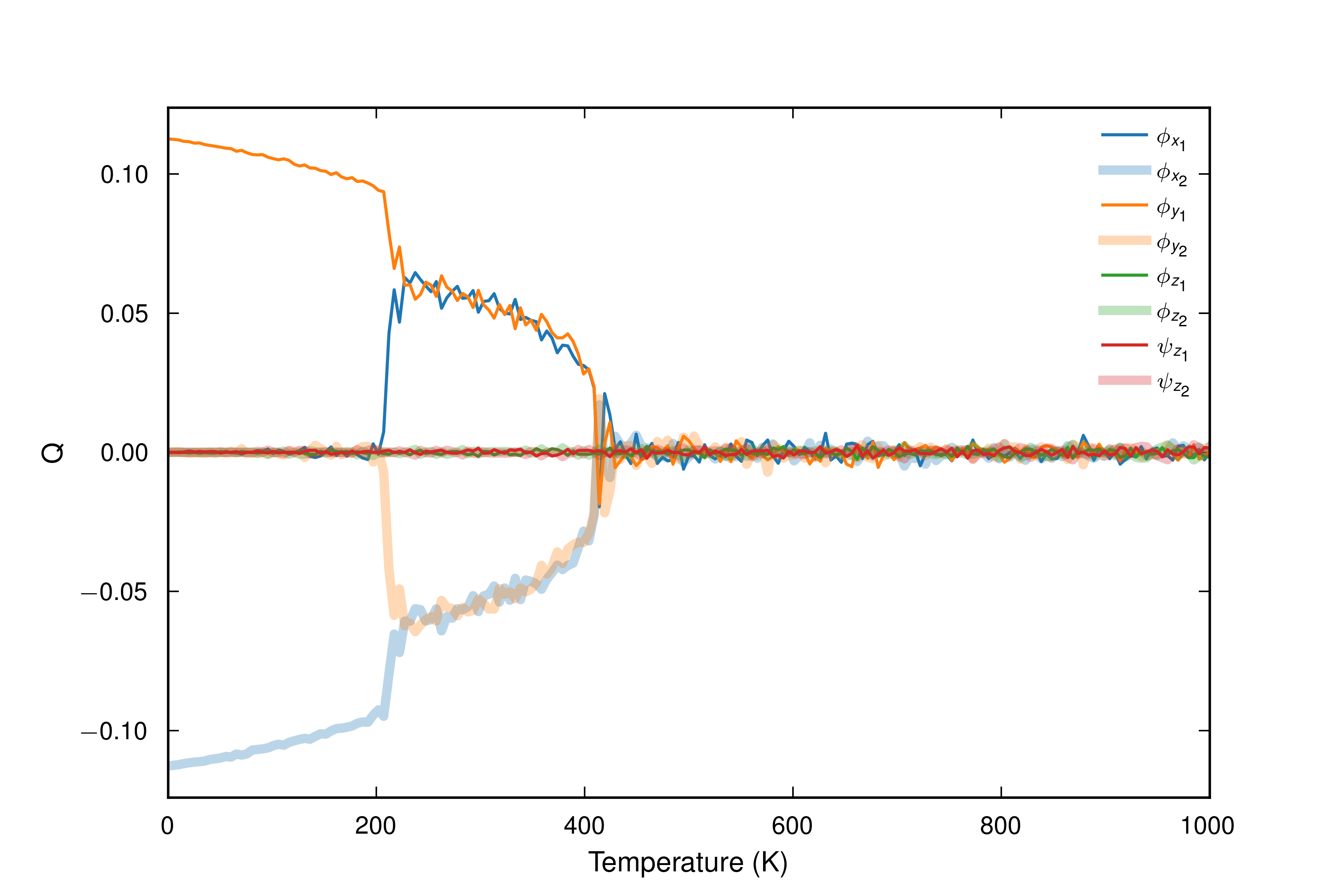}
    \caption{Mode projections in the cooling run for the $n = 1$ RP phase.}
    \label{fig:n1_cooling_tilt}
\end{figure}
\newpage
\begin{figure}[H]
    \centering
    \includegraphics[width=0.7\linewidth]{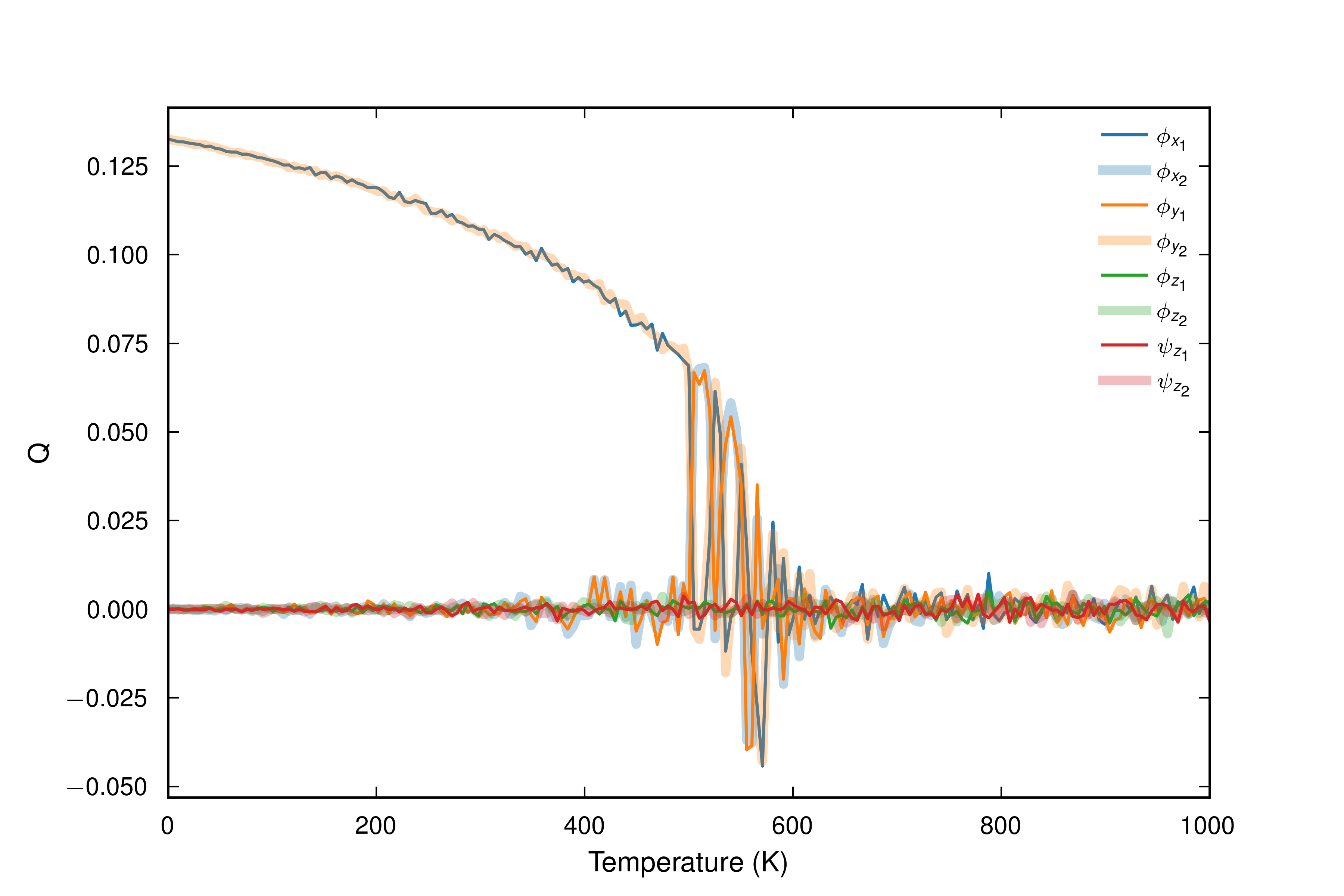}
    \caption{Mode projections in the heating run for the $n = 2$ RP phase.}
    \label{fig:n2_heating_tilt}
\end{figure}
\begin{figure}[H]
    \centering
    \includegraphics[width=0.7\linewidth]{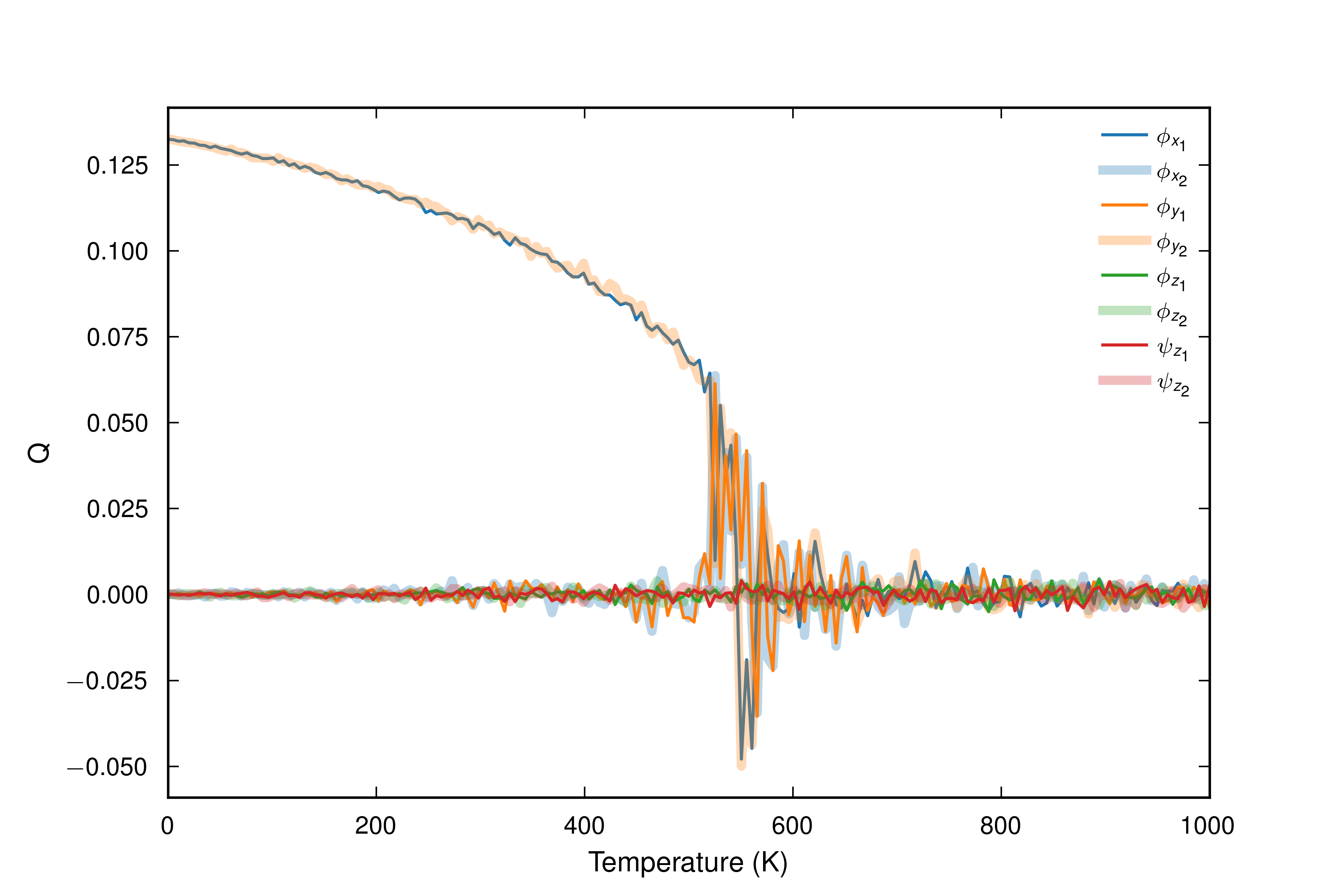}
    \caption{Mode projections in the cooling run for the $n = 2$ RP phase.}
    \label{fig:n2_cooling_tilt}
\end{figure}
\newpage
\begin{figure}[H]
    \centering
    \includegraphics[width=0.7\linewidth]{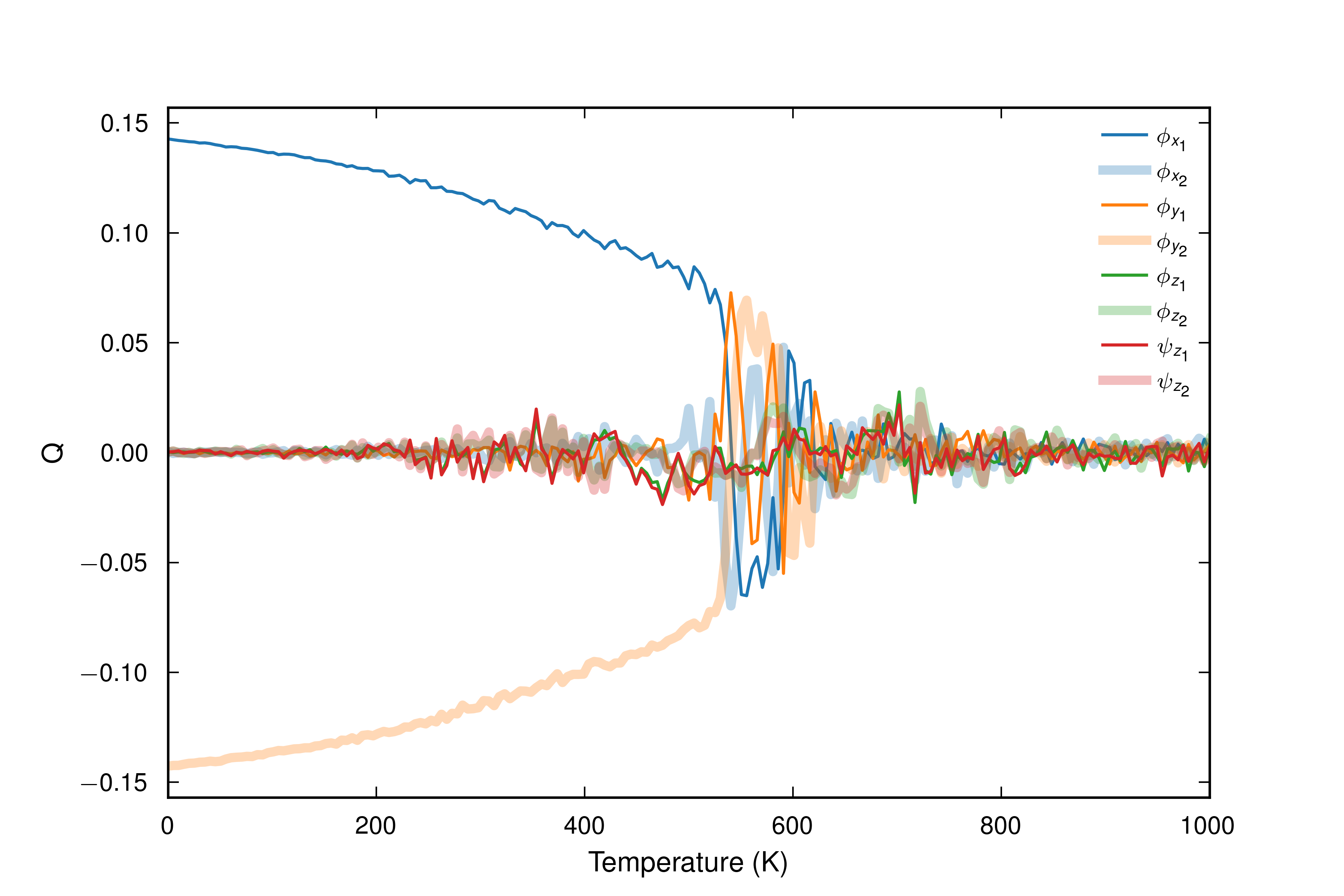}
    \caption{Mode projections in the heating run for the $n = 3$ RP phase.}
    \label{fig:n3_heating_tilt}
\end{figure}
\begin{figure}[H]
    \centering
    \includegraphics[width=0.7\linewidth]{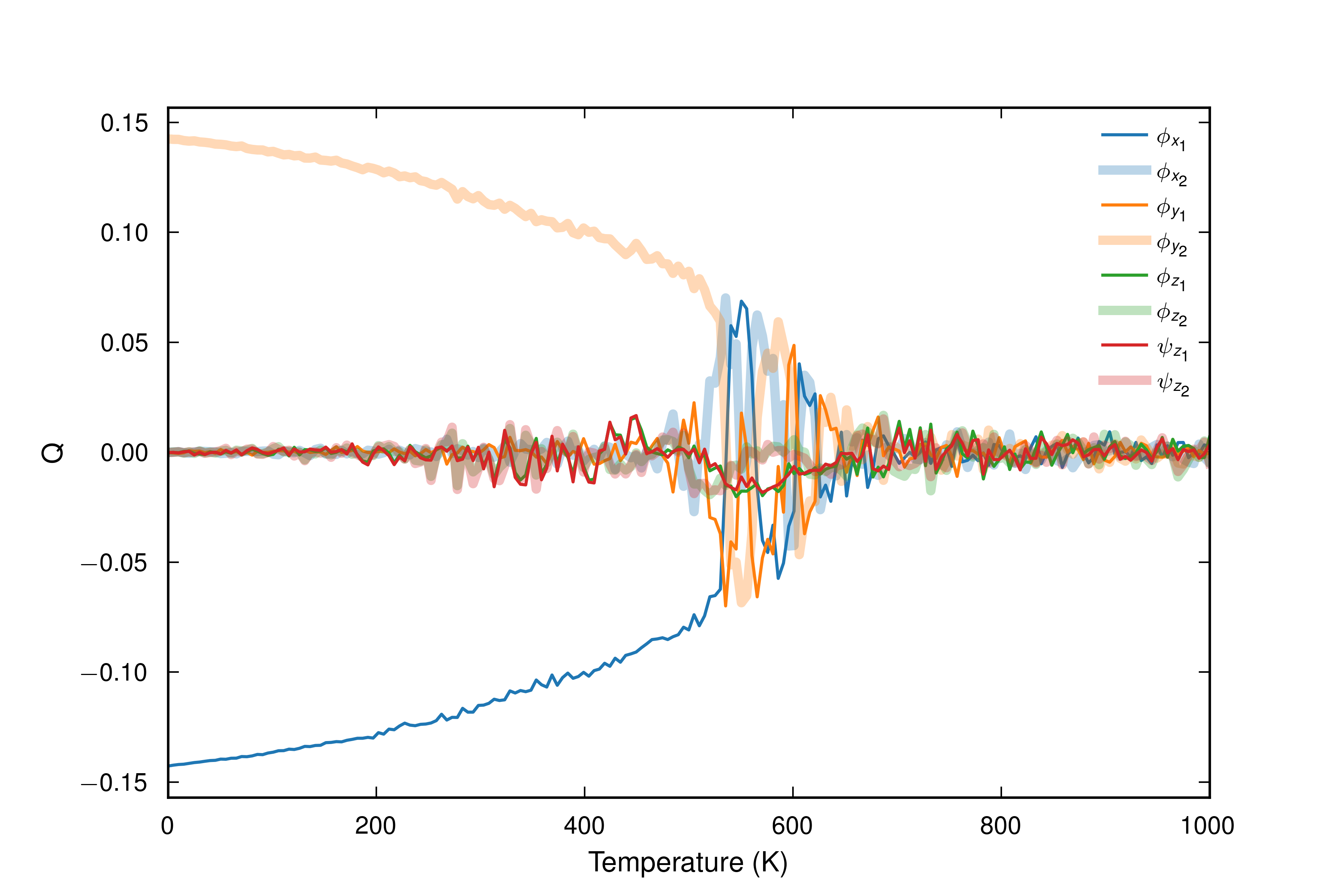}
    \caption{Mode projections in the cooling run for the $n = 3$ RP phase.}
    \label{fig:n3_cooling_tilt}
\end{figure}
\begin{figure}[H]
    \centering
    \includegraphics[width=0.7\linewidth]{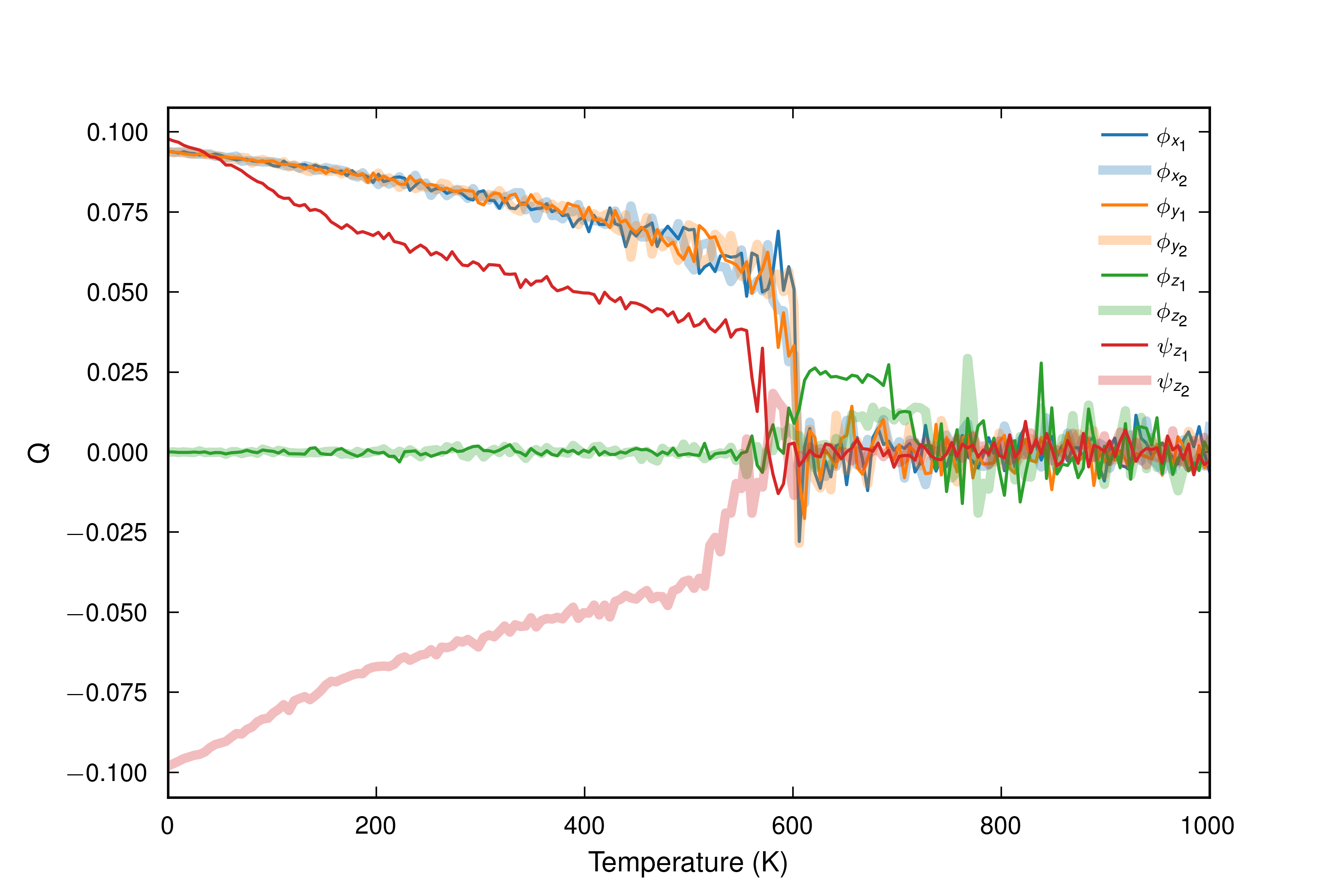}
    \caption{Mode projections in the heating run for the $n = 4$ RP phase.}
    \label{fig:enter-label}
\end{figure}
\begin{figure}[H]
    \centering
    \includegraphics[width=0.7\linewidth]{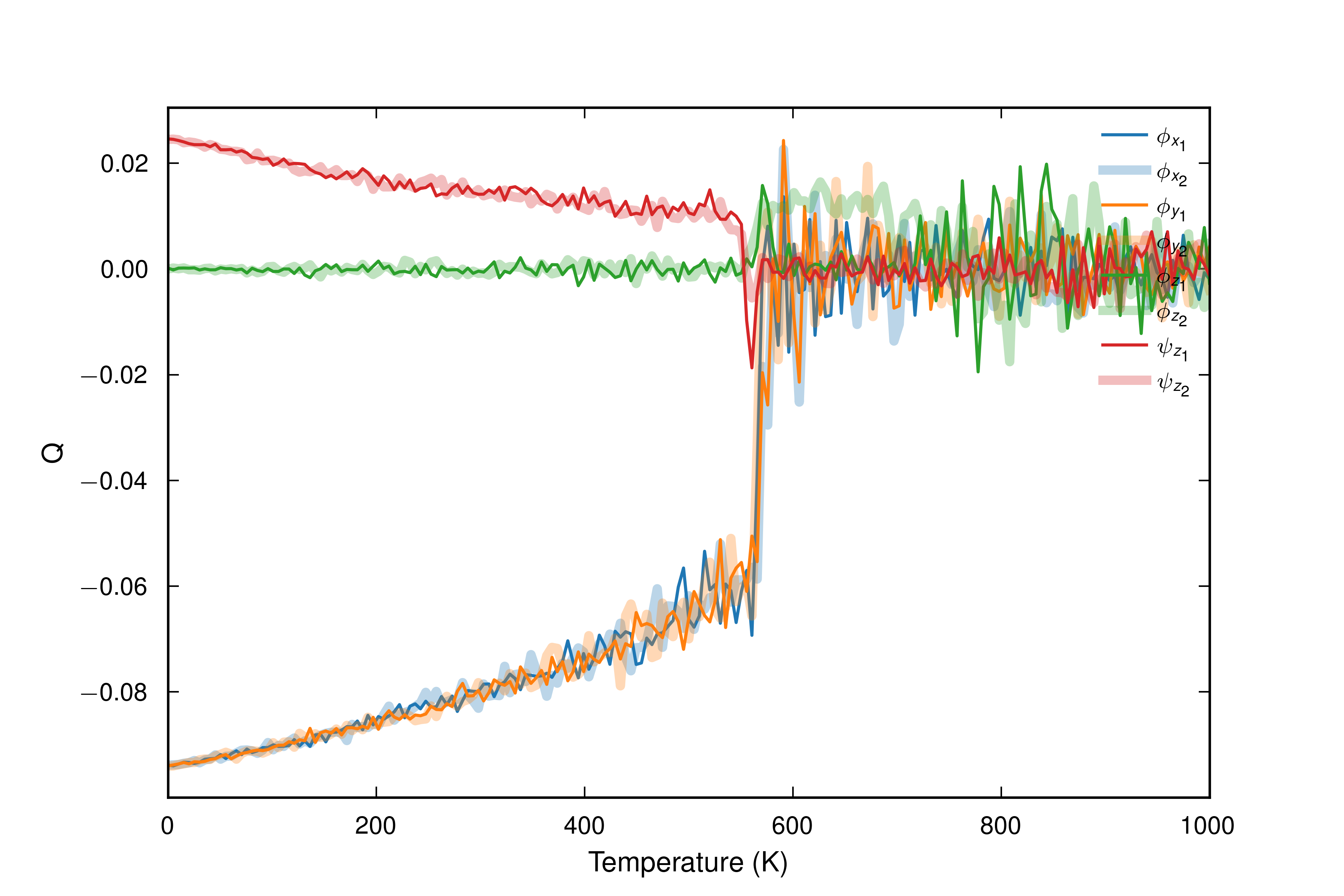}
    \caption{Mode projections in the cooling run for the $n=4$ RP phase.}
    \label{fig:enter-label}
\end{figure}
\begin{figure}[H]
    \centering
    \includegraphics[width=0.7\linewidth]{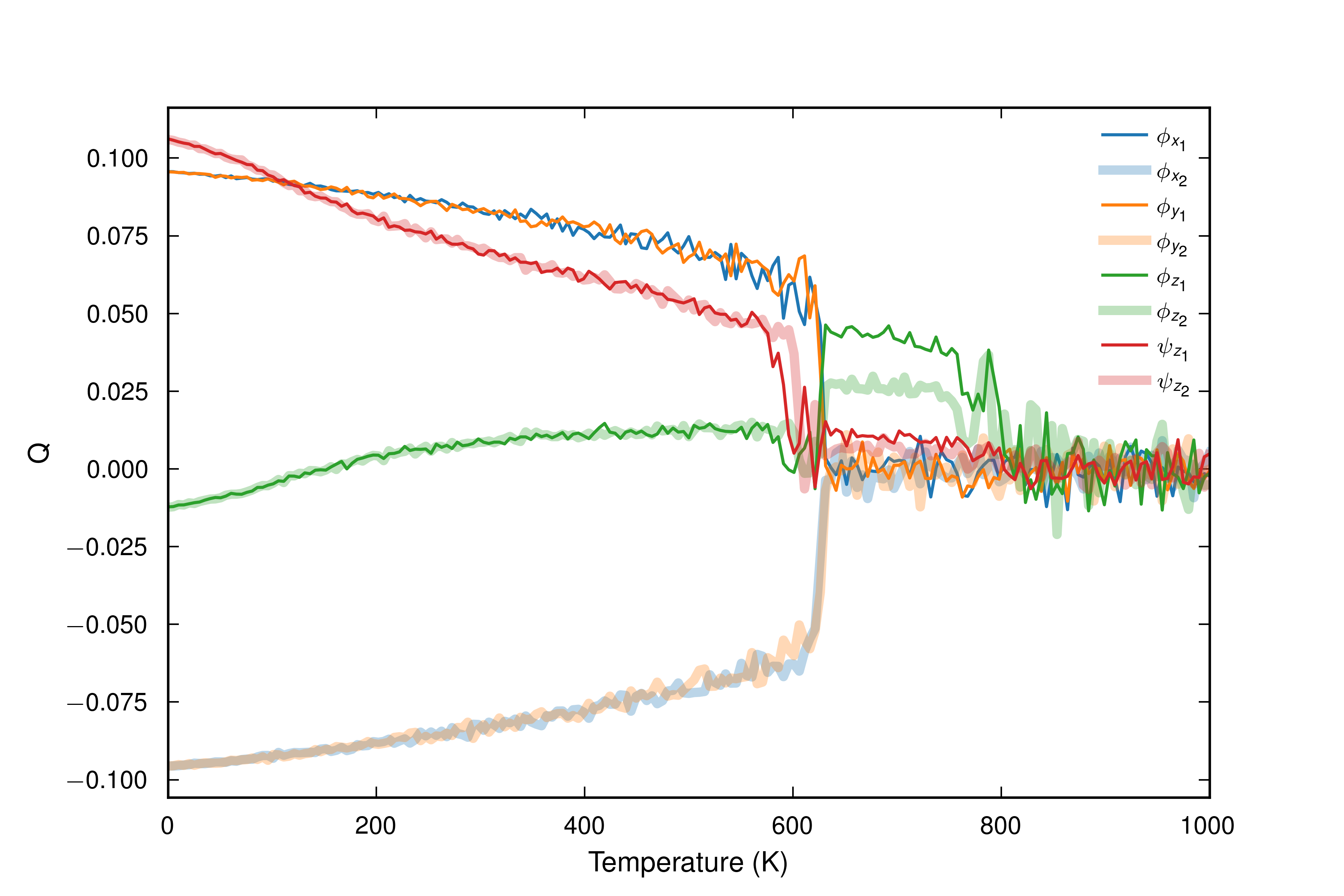}
    \caption{Mode projections in the heating run for the $n=5$ RP phase.}
    \label{fig:enter-label}
\end{figure}
\begin{figure}[H]
    \centering
    \includegraphics[width=0.7\linewidth]{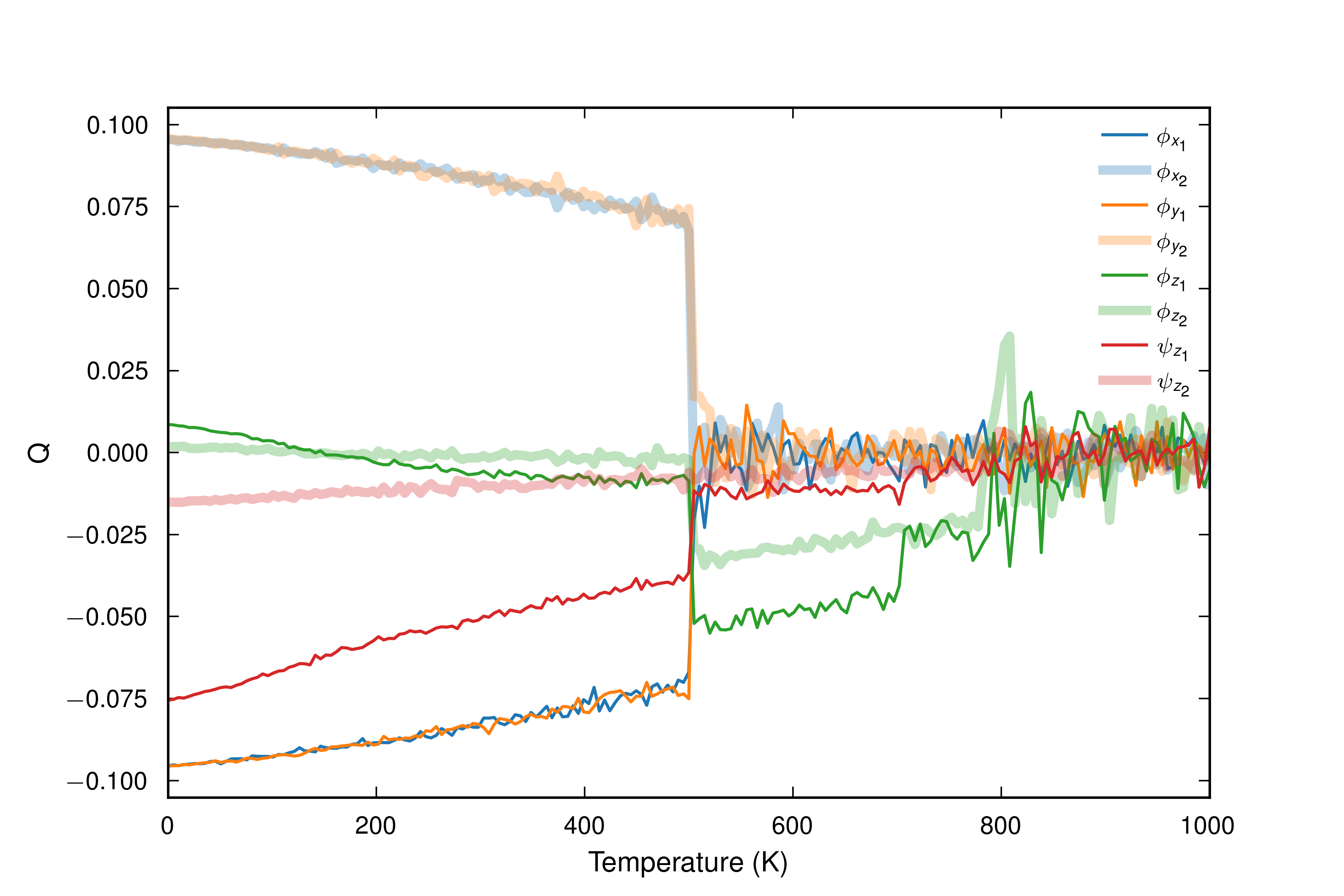}
    \caption{Mode projections in the cooling run for the $n=5$ RP phase.}
    \label{fig:enter-label}
\end{figure}
\begin{figure}[H]
    \centering
    \includegraphics[width=0.7\linewidth]{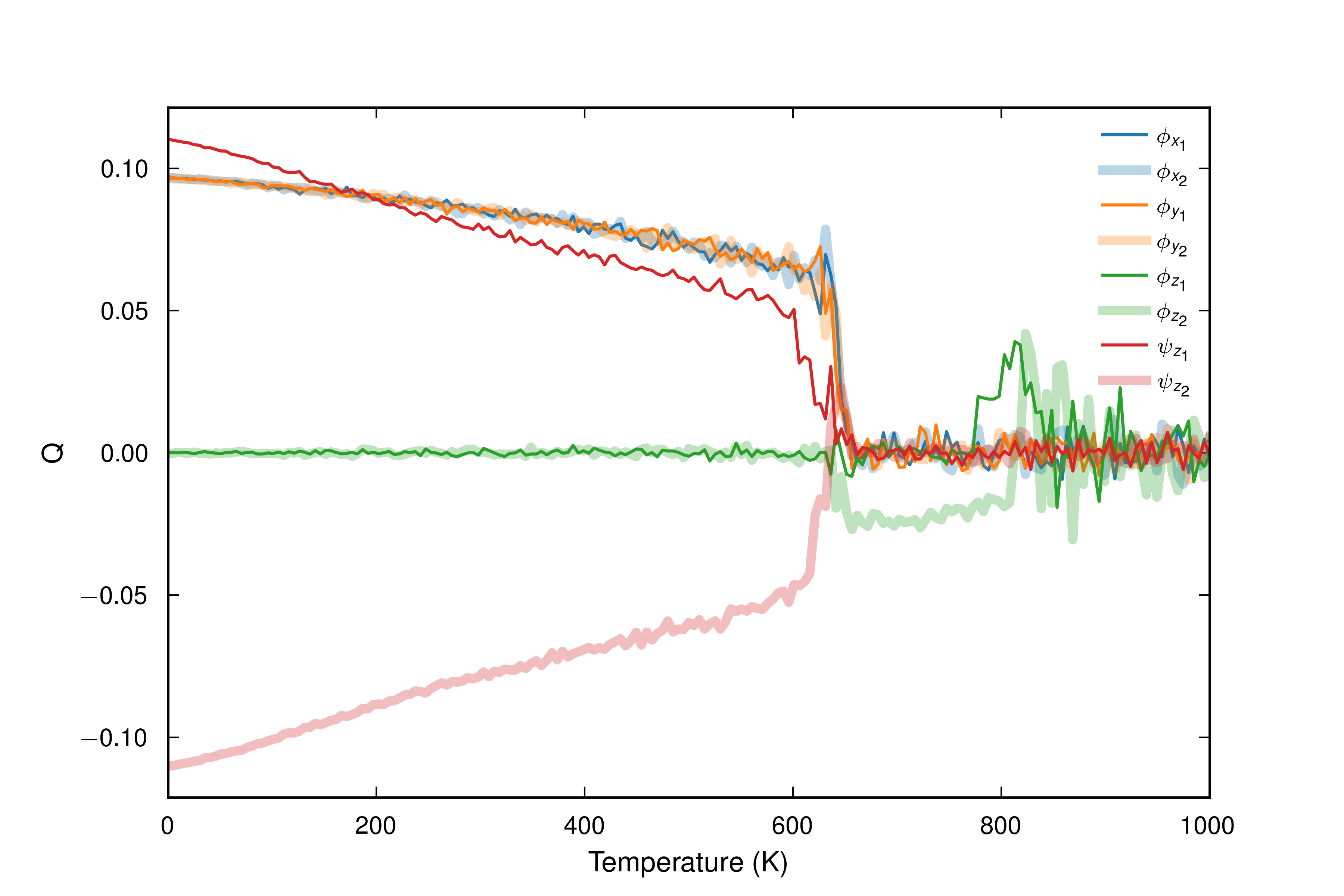}
    \caption{Mode projections in the heating run for the $n=6$ RP phase.}
    \label{fig:enter-label}
\end{figure}
\begin{figure}[H]
    \centering
    \includegraphics[width=0.7\linewidth]{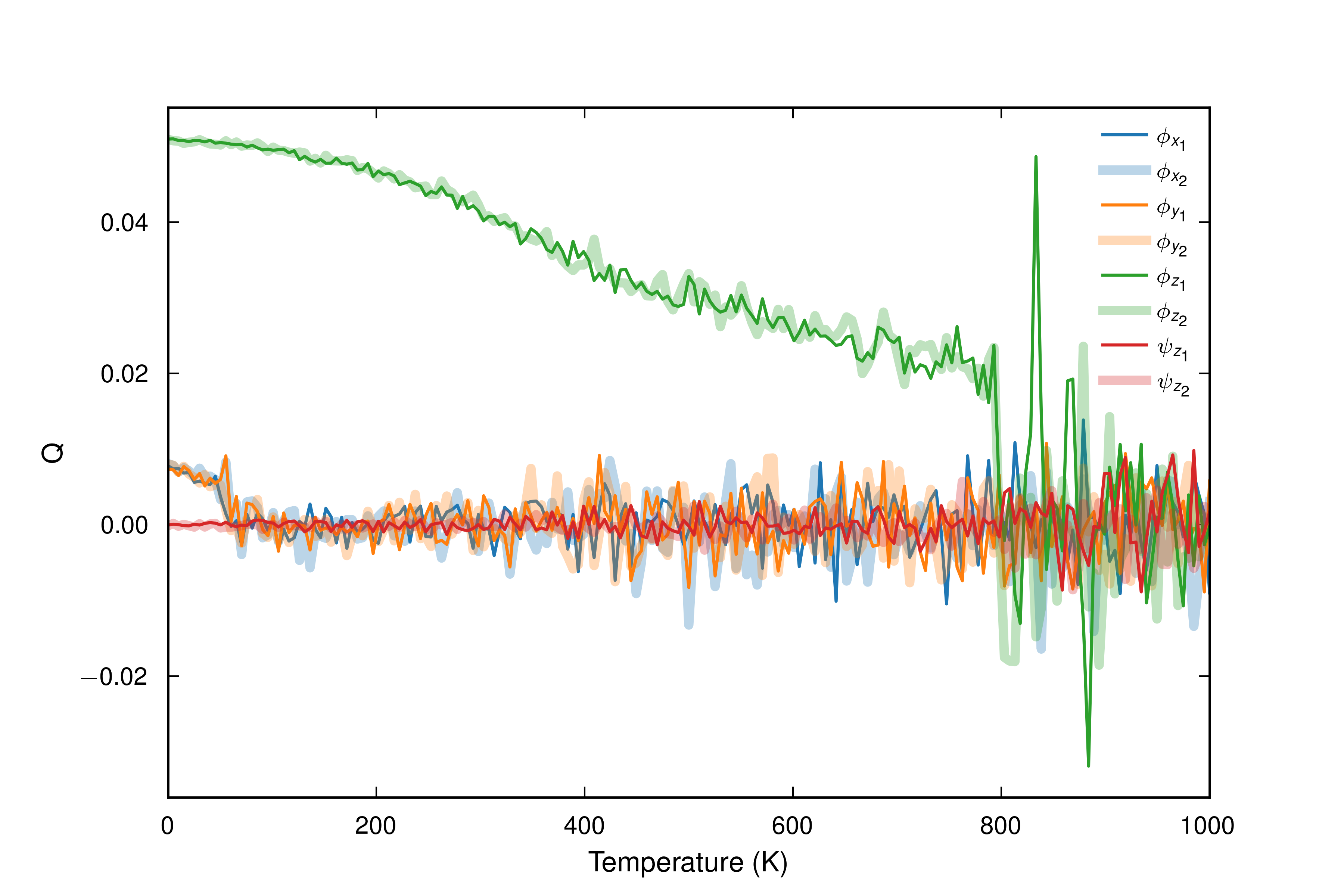}
    \caption{Mode projections in the cooling run for the $n=6$ RP phase.}
    \label{fig:enter-label}
\end{figure}

\newpage
\section{Tilt angle heatmaps}
Layer-by-layer Euler angles ($\theta_x$, $\theta_y$, $\theta_z$) of the \ce{ZrS6} octahedra in heating runs of the  $n = 1$ to $6$ RP phases.

In RP materials with $n\geq4$, the out-of-plane $\theta_z$ tilts at the perovskite-rocksalt interface (layer 1 and layer $n$) decrease as the rumpling amplitude increases (5b in the main text). The $\theta_z$ tilts vanish at the maximum of the rumpling amplitude.

\begin{figure}[H]
    \centering
    \includegraphics[width=0.5\linewidth]{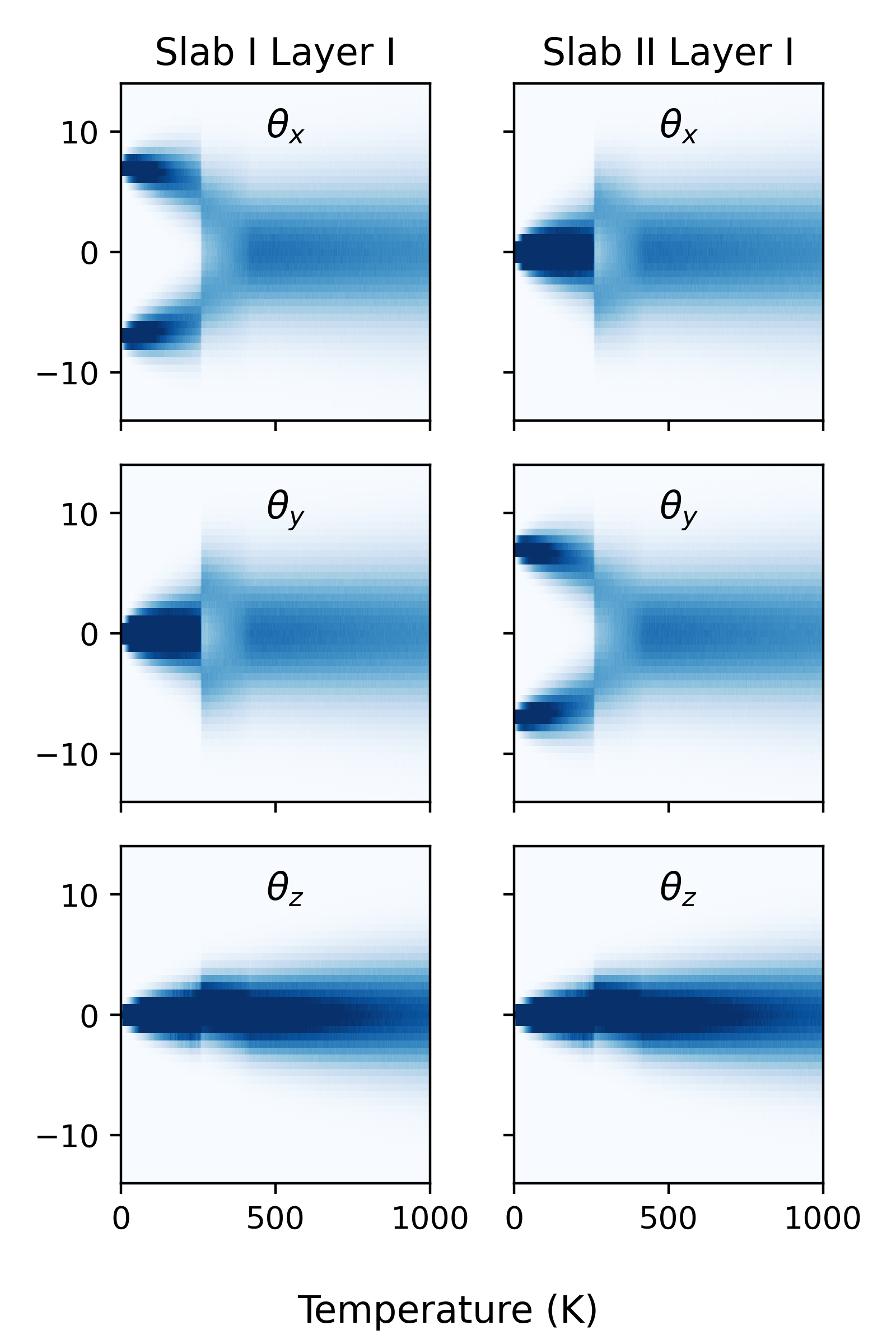}
    \caption{Euler angles of the $n=1$ RP phase}
    \label{fig:enter-label}
\end{figure}
\begin{figure}[H]
    \centering
    \includegraphics[width=0.5\linewidth]{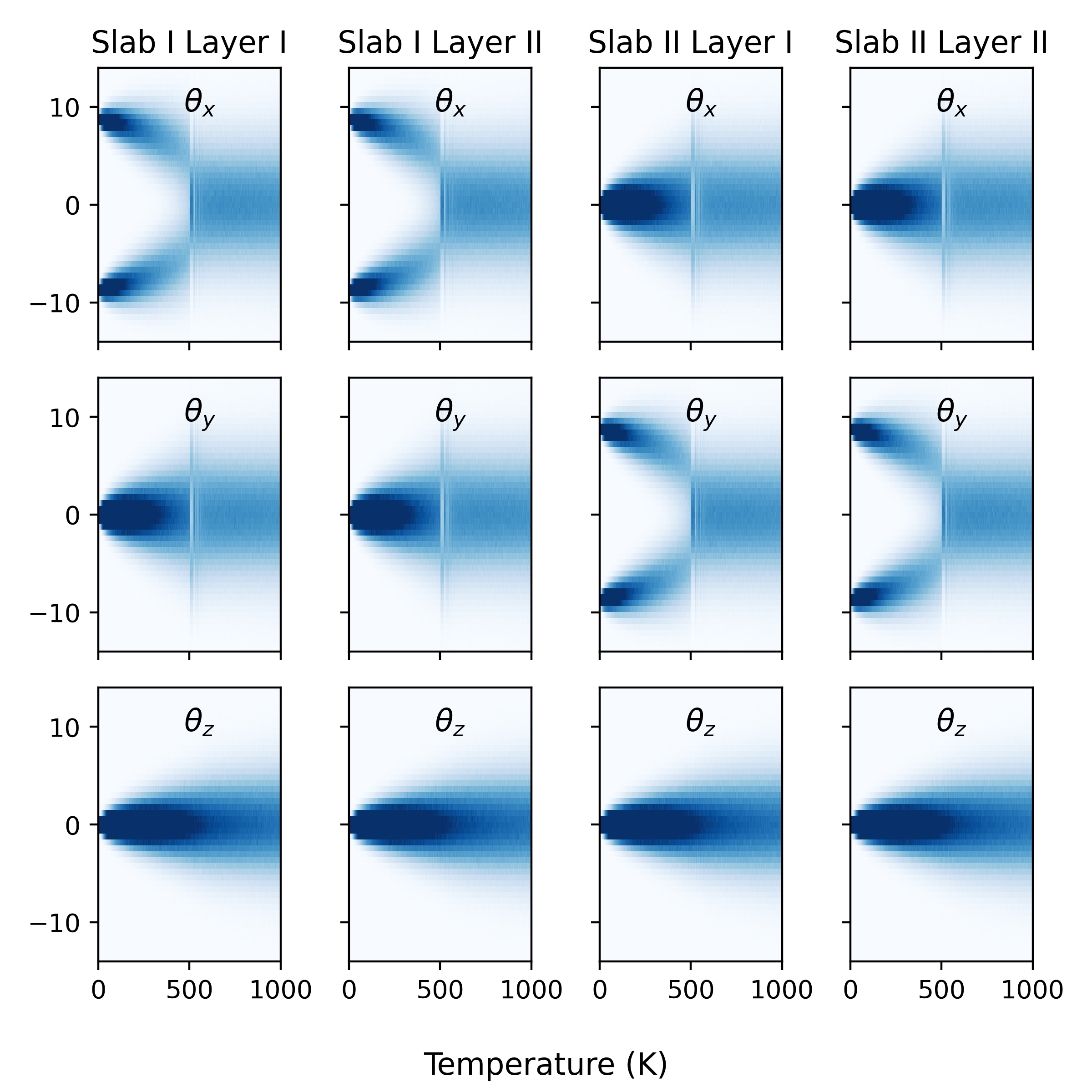}
    \caption{Euler angles of the $n=2$ RP phase}
    \label{fig:enter-label}
\end{figure}
\begin{figure}[H]
    \centering
    \includegraphics[width=\linewidth]{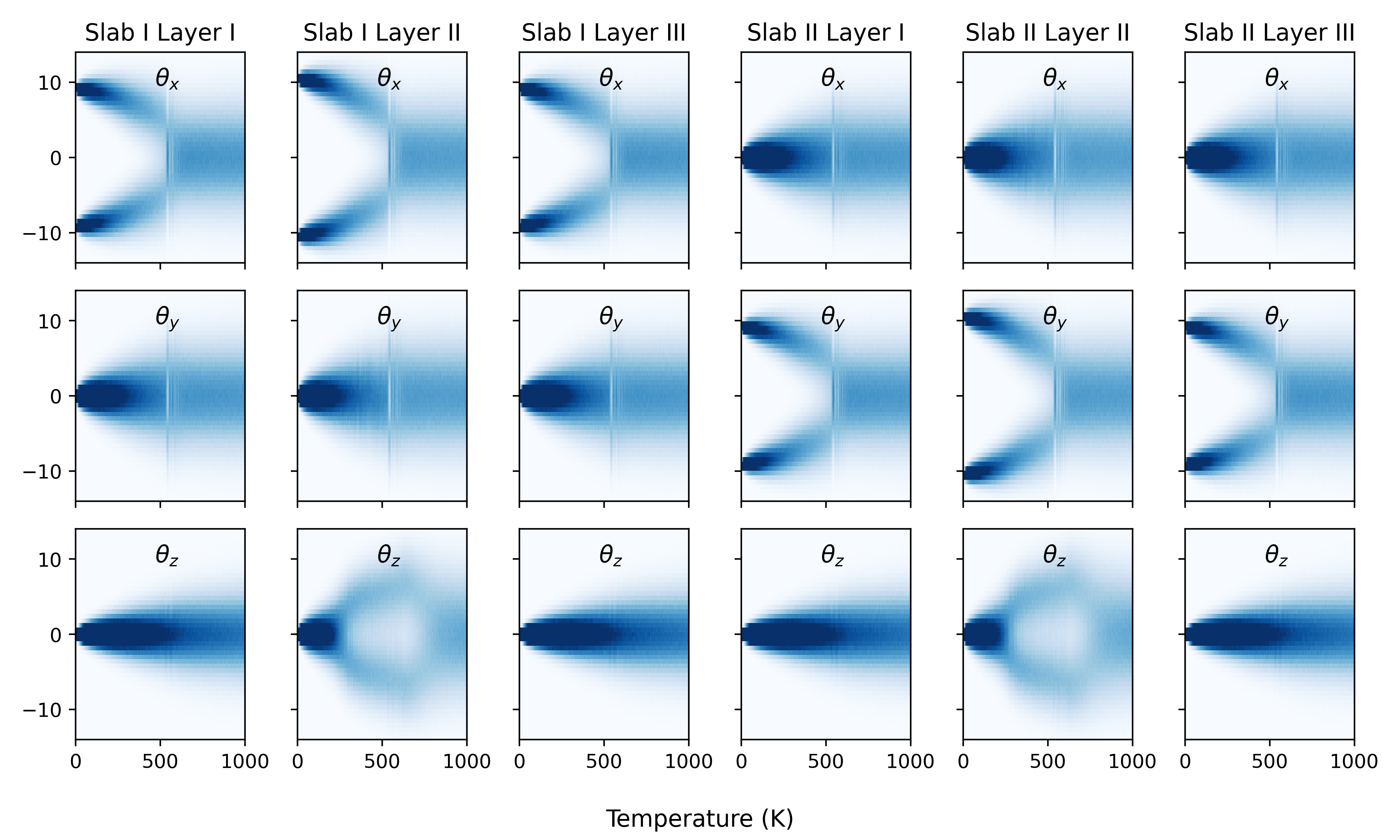}
    \caption{Euler angles of the $n=3$ RP phase}
    \label{fig:enter-label}
\end{figure}
\begin{figure}[H]
    \centering
    \includegraphics[width=\linewidth]{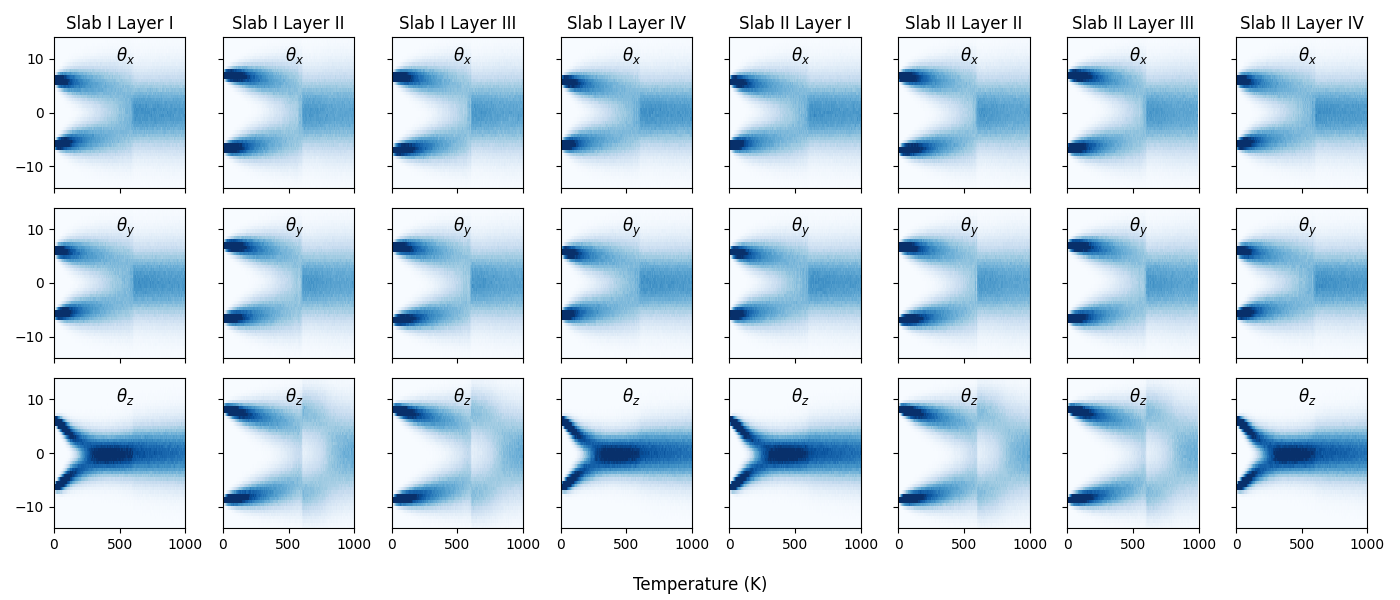}
    \caption{Euler angles of the $n=4$ RP phase}
    \label{fig:enter-label}
\end{figure}
\begin{figure}[H]
    \centering
    \includegraphics[width=\linewidth]{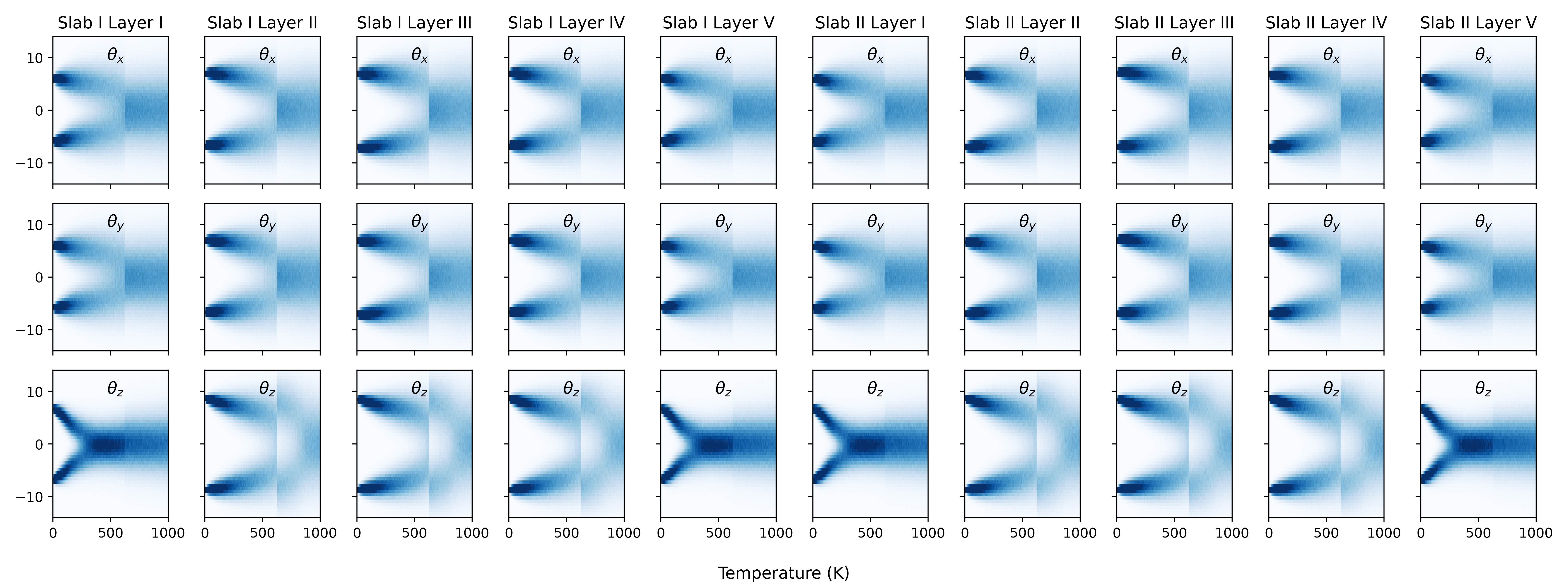}
    \caption{Euler angles of the $n=5$ RP phase}
    \label{fig:enter-label}
\end{figure}
\begin{figure}[H]
    \centering
    \includegraphics[width=\linewidth]{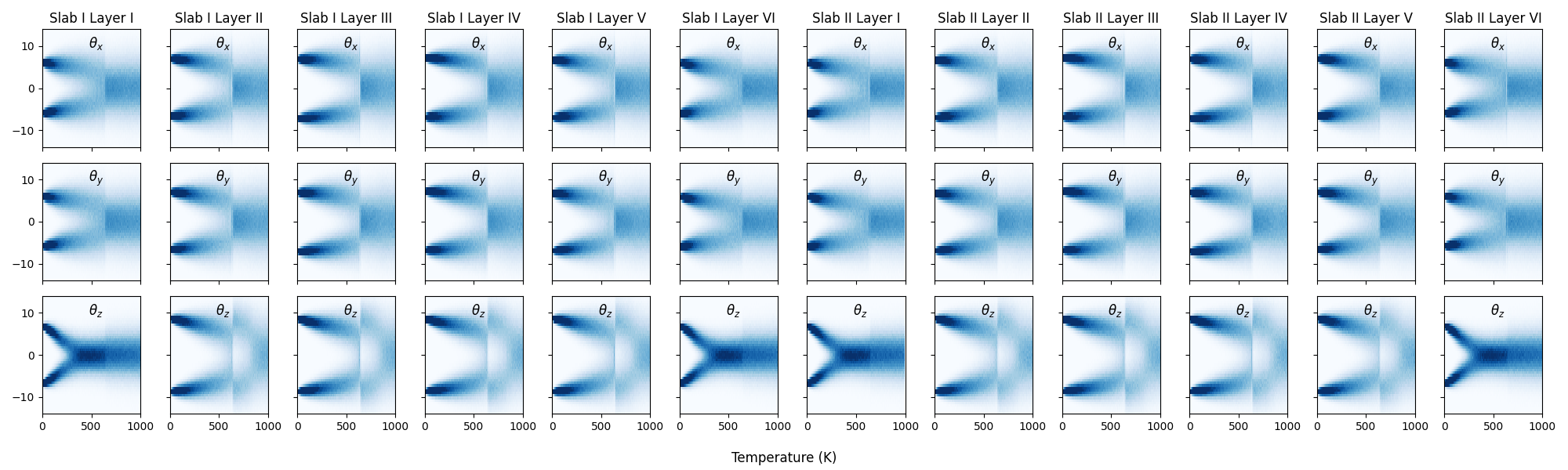}
    \caption{Euler angles of the $n=6$ RP phase}
    \label{fig:enter-label}
\end{figure}

\newpage

\section{Finite-temperature polymorphs}

Space group and tilt pattern for each of the polymorphs observed in the molecular dynamics simulations. 
Aleksandrov notation is used to denote the out-of-phase ($\phi$) and in-phase ($\psi$) octahedral tilts along each axis, with an overbar to indicate that the tilt patterns between adjacent slabs are in opposite directions \cite{aleksandrov1987successive}.
The $n=3$ \hmn{P2_1/c} phase corresponds to a more complex, layer-dependent tilt pattern (Fig. S24) in which there is a superposition of both $\phi_z$ and $\psi_z$ tilts (Fig. S10). This superposition is denoted with $\Delta$.

\begin{table}[H]
    \centering
    
    \begin{tabular}{lll}
        \toprule
        $n$ & Space group & Aleksandrov\\
        \midrule
        $n=1$ & \hmn{P4_2/ncm} & $\phi00 ~0\Bar{\phi}0$ \\
              & \hmn{Cmca} & $\phi\phi0 ~\Bar{\phi}\Bar{\phi}0$\\
              & \hmn{I4/mmm} & $000 ~000$\\
        $n=2$ & \hmn{P4_2/mnm} & $\phi00 ~0\phi0$\\
              & \hmn{I4/mmm} & $000 ~000$\\
        $n=3$ & \hmn{P4_2/ncm} & $\phi0 ~0\Bar{\phi}0$\\
              & \hmn{P2_1/c} & $\phi0\Delta ~0\Bar{\phi}\Delta$\\
              & \hmn{Cmca} & $00\phi_z ~00\phi_z$\\
              & \hmn{I4/mmm} & $000 ~000$\\
        $n=4$ & \hmn{Pnma} & $\phi\phi\psi_z ~\phi\phi\Bar{\psi}_z$\\
              & \hmn{Ccca} & $00\phi_z ~00\phi_z$\\
              & \hmn{I4/mmm} & $000 ~000$\\
        $n=5$ & \hmn{P2_1/c} & $\phi\phi\psi_z ~\Bar{\phi}\Bar{\phi}\Bar{\psi}_z$\\
              & \hmn{Cmca} & $00\phi_z ~00\phi_z$\\
              & \hmn{I4/mmm} & $000 ~000$\\
        $n=6$ & \hmn{Pnma} & $\phi\phi\psi_z ~\phi\phi\Bar{\psi}_z$\\
              & \hmn{Ccca} & $00\phi_z ~00\phi_z$\\
              & \hmn{I4/mmm} & $000 ~000$\\
        $n=\infty$ & \hmn{Pnma} & $\phi\phi\psi_z$\\
                   & \hmn{I4/mcm} & $00\phi_z$\\
                   & \hmn{Pm-3m} & $000$\\
    \bottomrule
    \end{tabular}
    \caption{Space group and tilt pattern for each of the polymorphs observed in this study.}
    \label{tab:tilt_patterns}
\end{table}

\clearpage

\section{Octahedral tilt angles for $n=1$ to $n=4$ RP phases}
\begin{figure}[H]
    \centering
    \includegraphics[width=\linewidth]{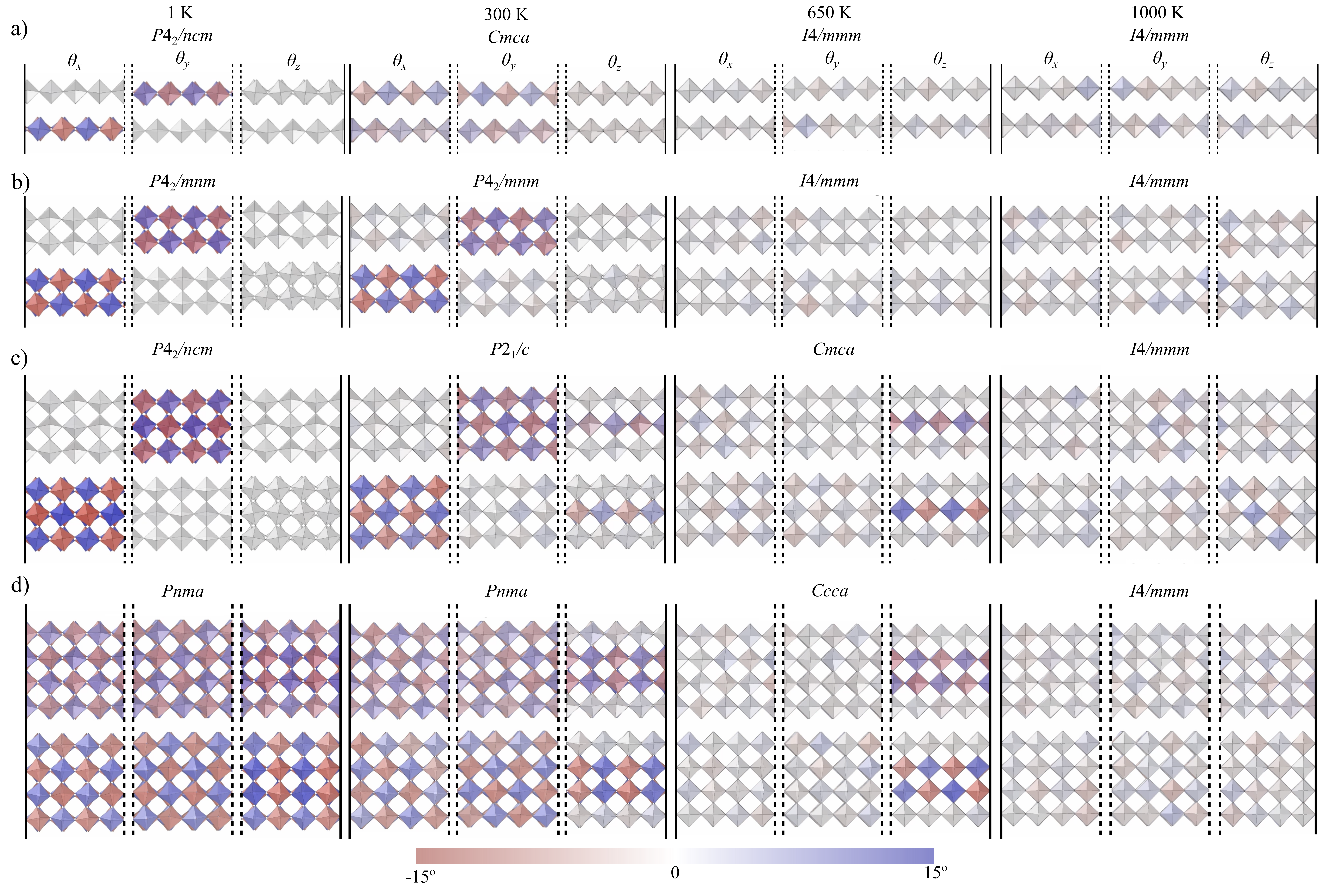}
    \caption{Snapshots of Euler angles ($\theta_x$, $\theta_y$, $\theta_z$) of the \ce{ZrS6} octahedra in the a) $n=1$, b) $n=2$, c) $n=3$ and d) $n=4$ RP polymorphs at four temperature points. The octahedra are colour-coded according to the magnitude of their tilt angles. 
    Solid lines separate the structures by temperature, dashed lines separate the structures by tilt axis. The structures are viewed along the x-axis for $\theta_x$ and $\theta_z$, and along the y-axis for $\theta_y$.}
    \label{fig:enter-label}
\end{figure}

\newpage

\section{A-site displacements}
Rumpling refers to the displacement of A-site cations along the $z$-direction. Here, we present the layer-resolved rumpling amplitudes for RP phases with $n=1$ to 6.
Our analysis reveals that rumpling is most pronounced at the outermost layers, at the perovskite-rocksalt interfaces.
The rumpling amplitudes are symmetric across the layers of the slab and remain the same between the two slabs in each RP phase.
\begin{figure}[H]
    \centering
    \includegraphics[width=0.8\linewidth]{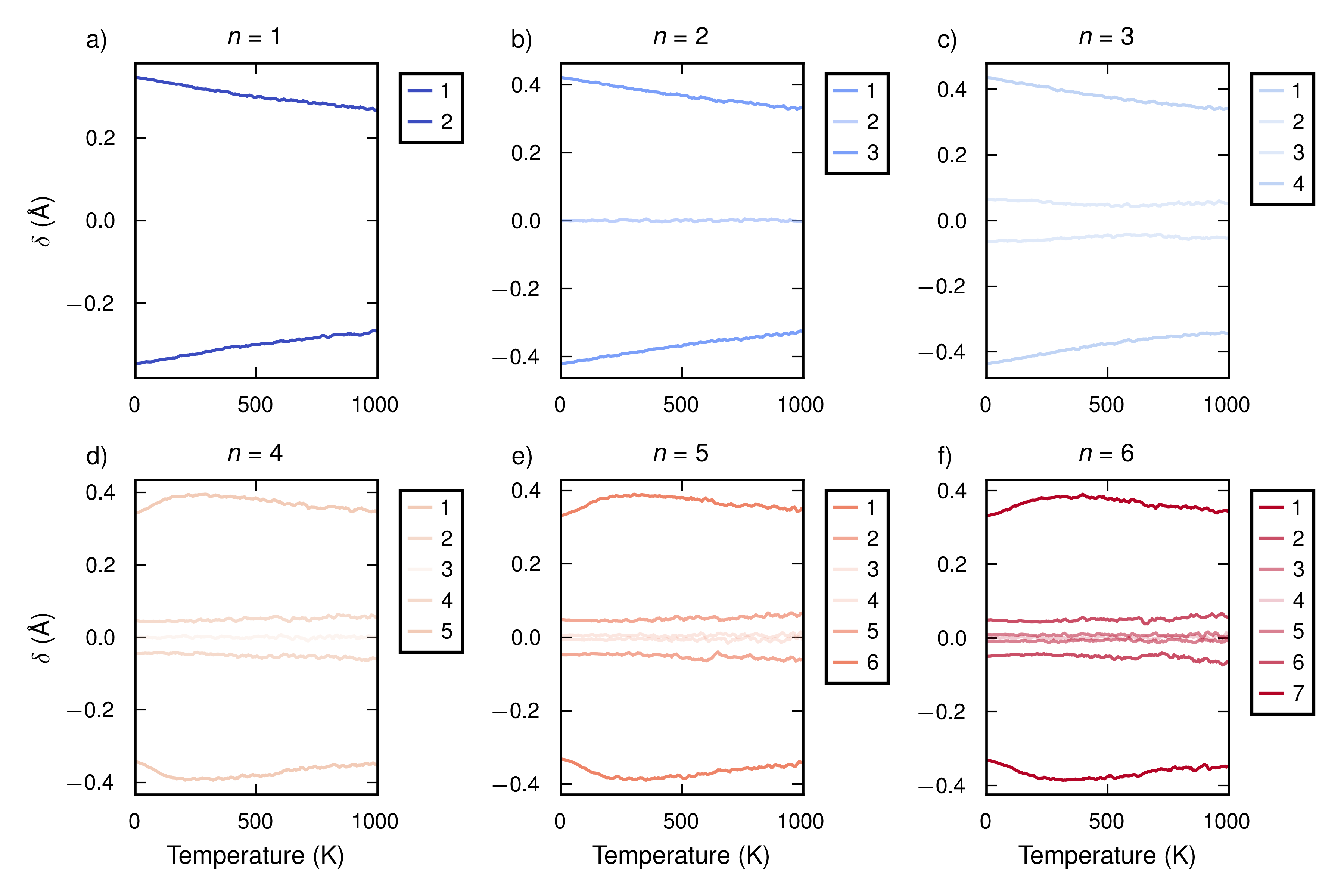}
    \caption{Layer-by-layer rumpling amplitude in the $n=1$ to 6 RP phases}
    \label{fig:enter-label}
\end{figure}

As has been observed in previous publications, the competition between rumpling and out-of-plane octahedral tilting results from both distortions reducing the cation-anion distance (in this case between Ba and S) \cite{Zhang2020unraveling}. This is demonstrated in \autoref{tab:distort}, where the distance between Ba and S atoms at the perovskite-rocksalt interface are provided for the untilted \hmn{I4/mmm} structures and the ground-state tilted structure. In both cases a rumpling distortion has also been applied.

\begin{table}[H]
\centering
\begin{tabular}{lll}
\toprule
$n$ &\hmn{I4/mmm} &Ground state \\
\midrule
1&3.56& 3.26 \\ 
2&3.58&3.24 \\ 
3&3.59&3.24 \\ 
4&3.59&3.21\\ 
5&3.59&3.21 \\ 
6&3.59&3.21 \\ 
\bottomrule
\end{tabular}
\caption{In-plane Ba-S distances in \ce{Ba_{n+1}Zr_{n}S_{3n+1}} for the untilted \hmn{I4/mmm} structures and the ground-state tilted structure. Distances are given in \SI{}{\angstrom}. In both cases a rumpling distortion has been applied.}
\label{tab:distort}
\end{table}

\begin{figure}[H]
    \centering
    \includegraphics[width=0.49\linewidth]{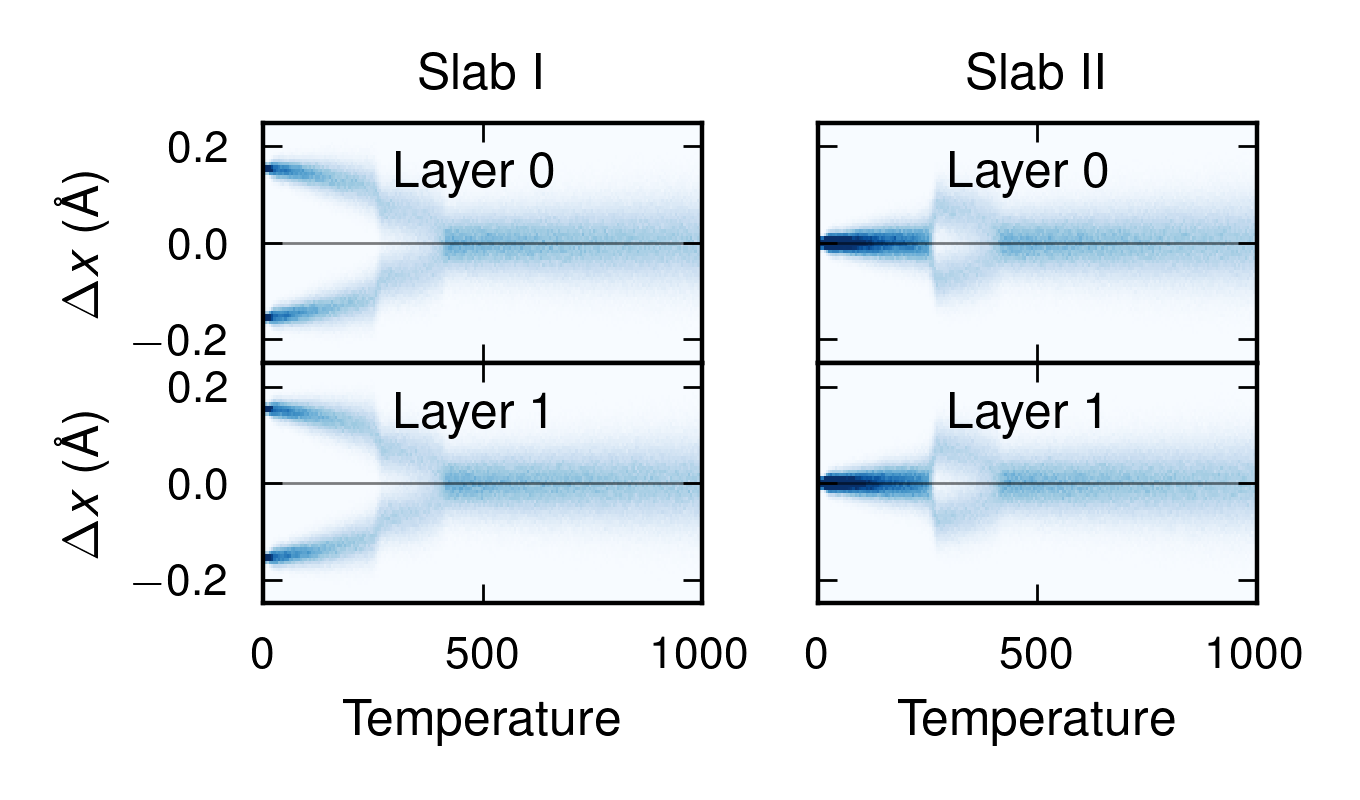}
    \includegraphics[width=0.49\linewidth]{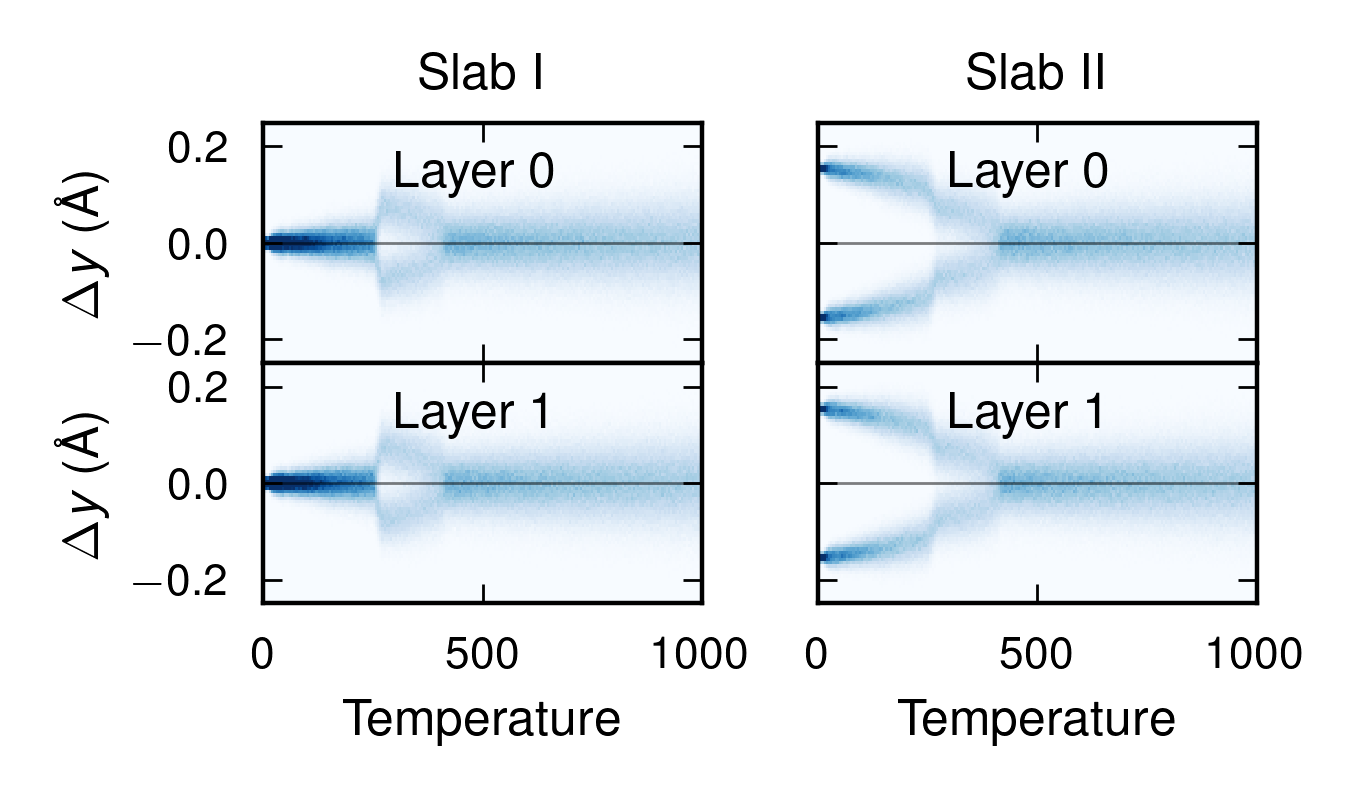}
    \caption{A-site displacements in the $n=1$ RP phase}
    \label{sfig:A_site_xy_n1}
\end{figure}

\begin{figure}[H]
    \centering
    \includegraphics[width=0.49\linewidth]{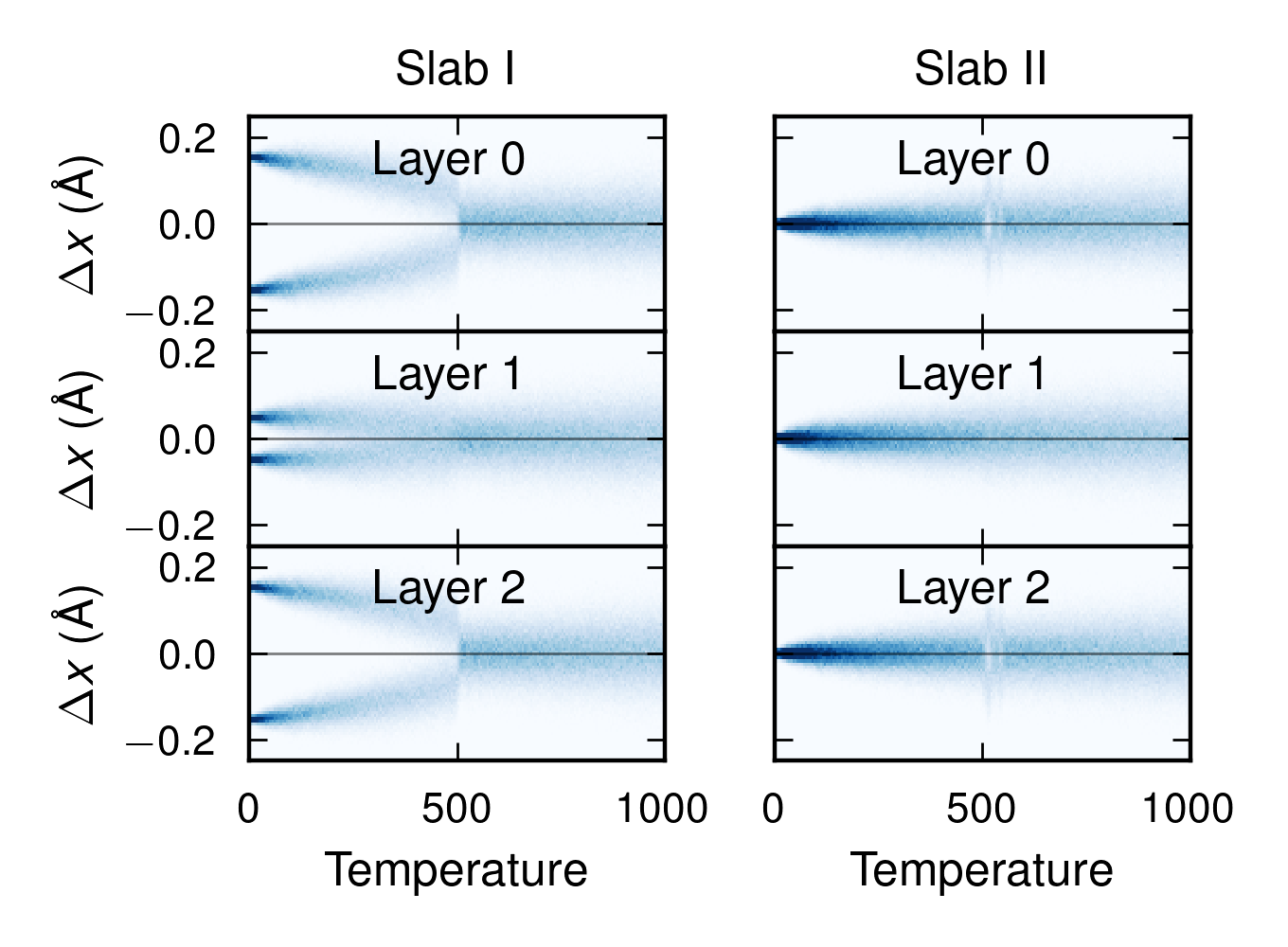}
    \includegraphics[width=0.49\linewidth]{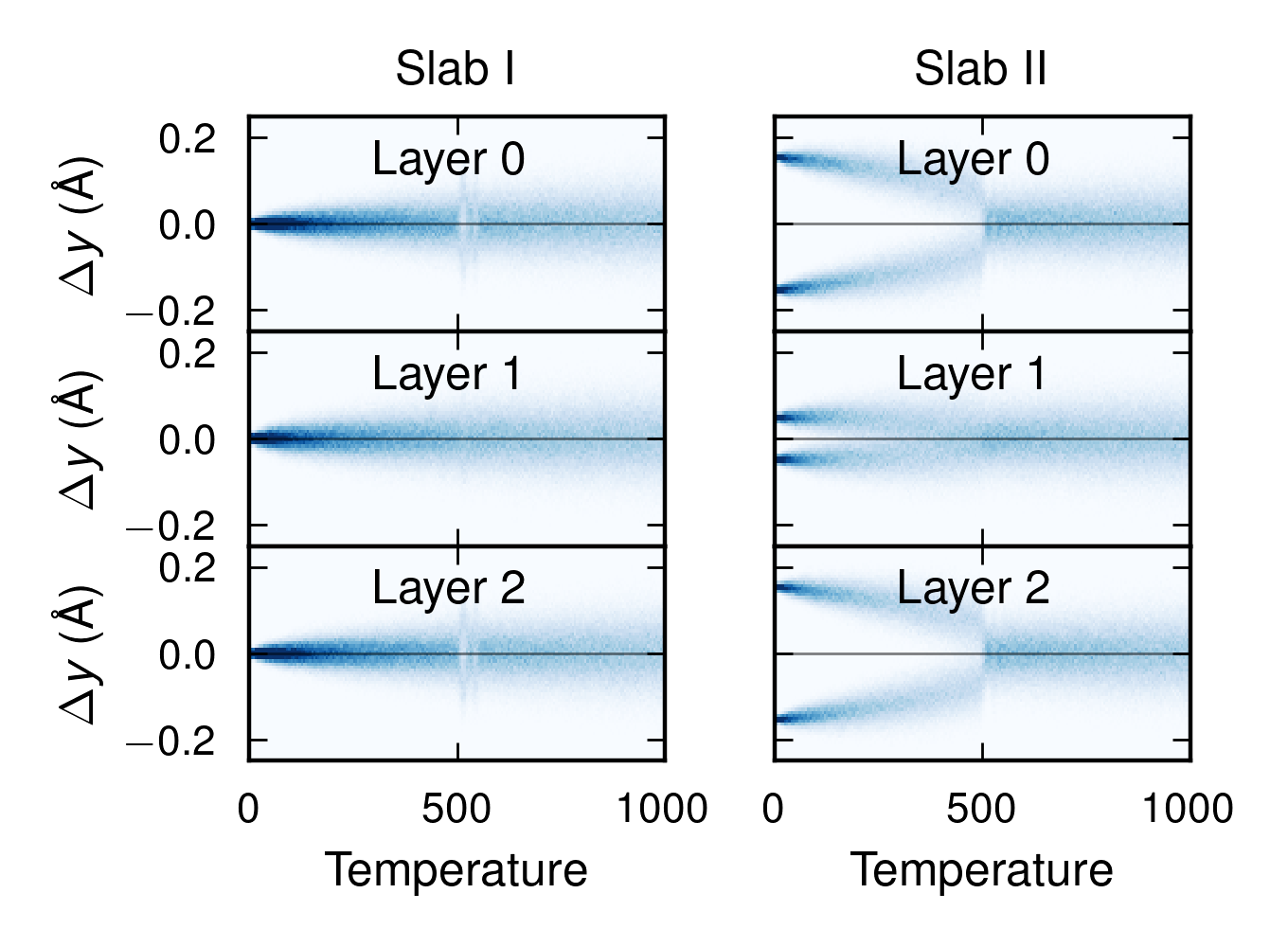}
    \caption{A-site displacements in the $n=2$ RP phase}
    \label{sfig:A_site_xy_n2}
\end{figure}

\begin{figure}[H]
    \centering
    \includegraphics[width=0.49\linewidth]{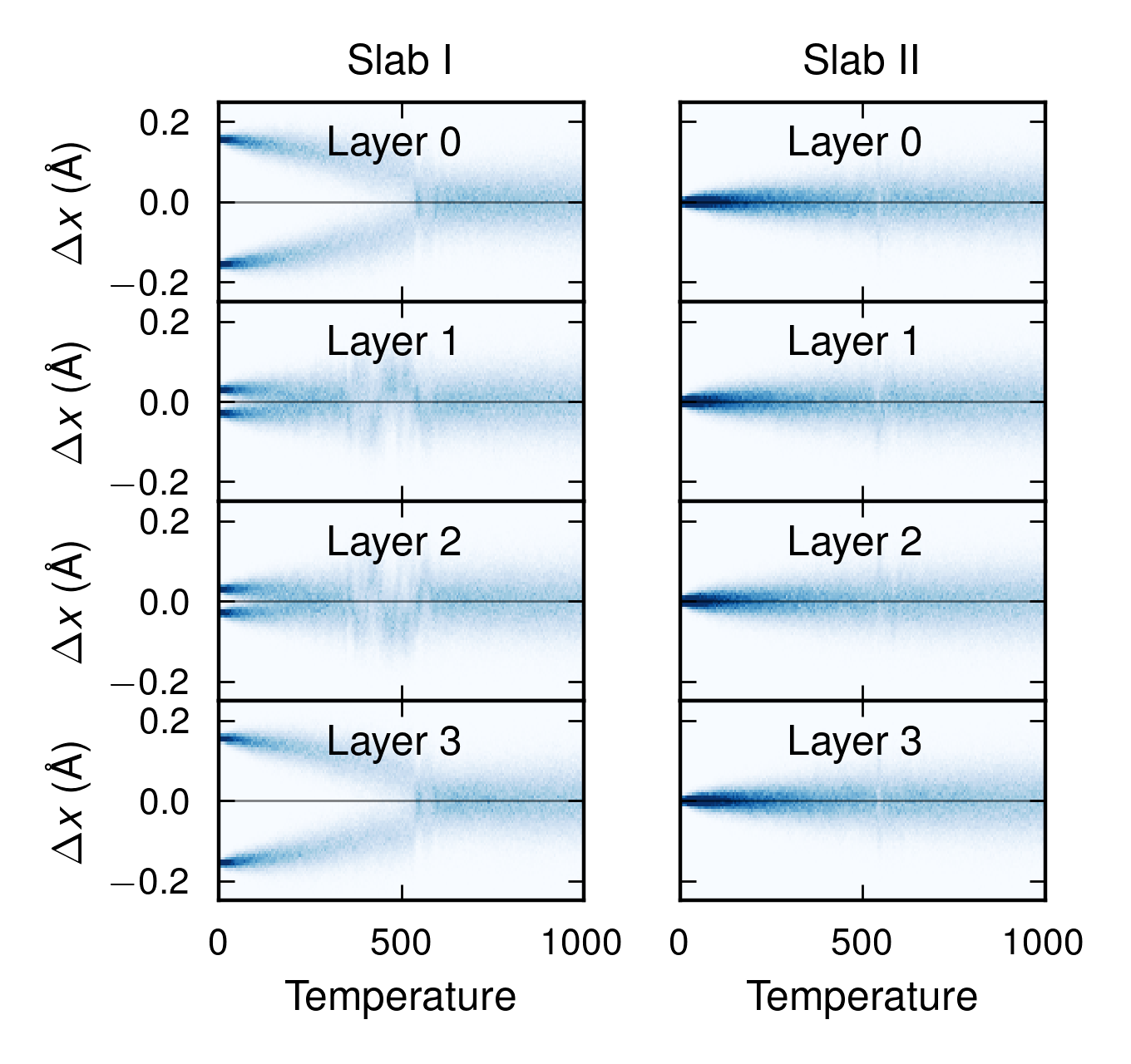}
    \includegraphics[width=0.49\linewidth]{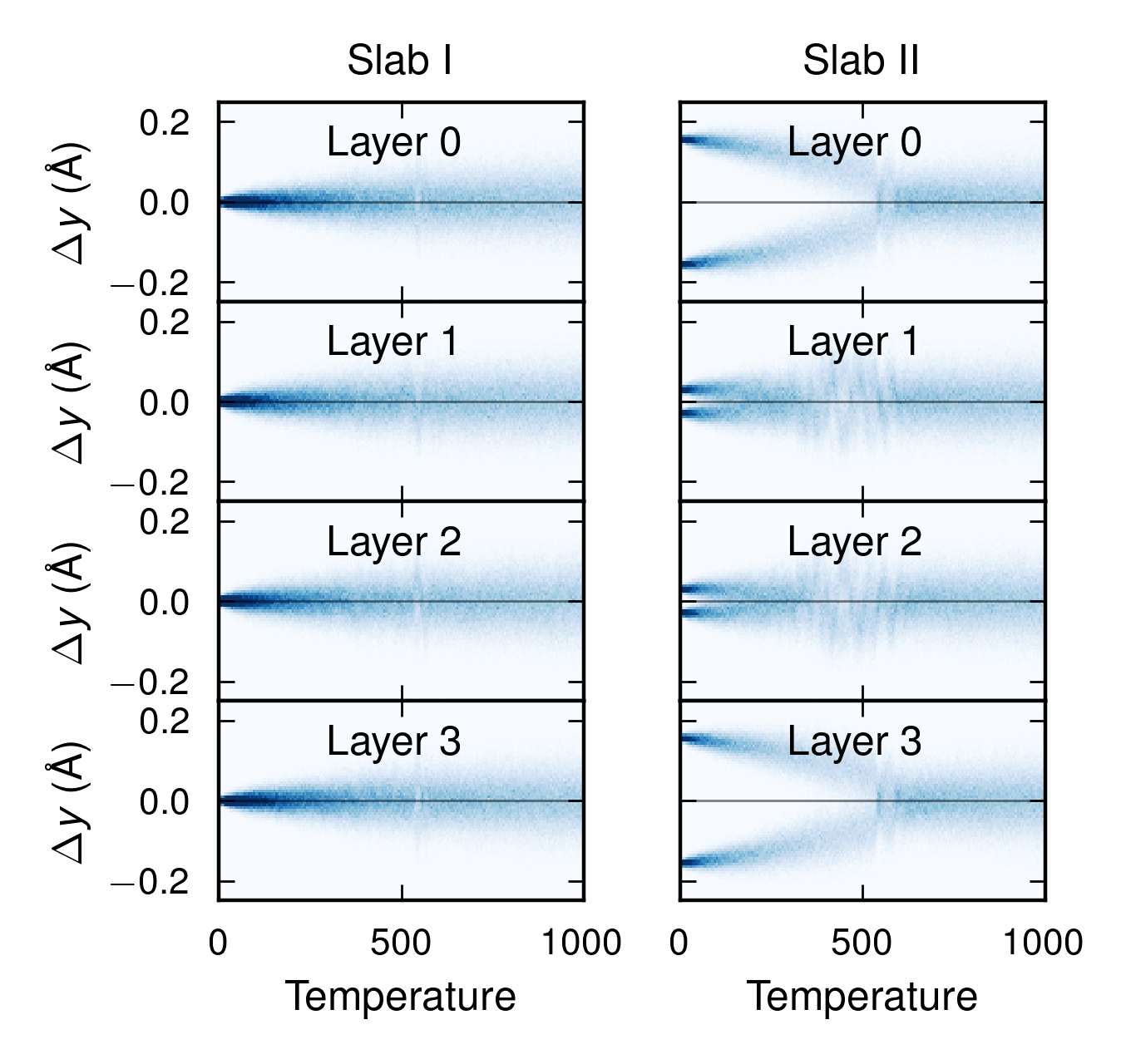}
    \caption{A-site displacements in the $n=3$ RP phase}
    \label{sfig:A_site_xy_n3}
\end{figure}

Next, we look at A-site displacements in the $xy$-plane.
Here, the displacements are computed with the high-symmetry \hmn{I4/mmm} structure as reference.
For $n=1,2,3$ the A-site displacements alternate within a layer (normal along z-axis) - see \autoref{sfig:A_site_xy_n1}, \autoref{sfig:A_site_xy_n2}, \autoref{sfig:A_site_xy_n3}.
Additionally, one can clearly see how the A-site displacements in the $x$ and $y$ directions follows that of the out of phase tilting around the $x$ and $y$ axes.
For example, in $n=1$ at \qty{0}{\kelvin} we have a tilting pattern of $\phi00 ~ 0\bar{\phi}0$, and A-sites in the first slab are only displaced in the x-direction whereas in second slab they are displaced in the y-direction.
As the system transitions to $\phi\phi0 ~ \bar{\phi}\bar{\phi}0$ we see both slabs exhibiting A-site displacements along both $x$ and $y$ direction.
Finally, as the system transitions to untilted \hmn{I4/mmm} phase the A-site displacements disappear.

In $n=4$ (and above) the in-plane A-site displacements within each layer are activated in $x$ and $y$ directions. This follows from the tilt pattern $\phi\phi\psi_z ~ \phi\phi\bar{\psi}_z$.
We note that within slab I there is a net displacement of A-sites, however this cancels out as the net displacements in slab II is in the opposite direction.
This is expected for the \hmn{Pnma} structure as it does not exhibit the improper ferroelectric behaviour seen in the \hmn{Cmc2_1} structure.

\begin{figure}[H]
    \centering
    \includegraphics[width=0.49\linewidth]{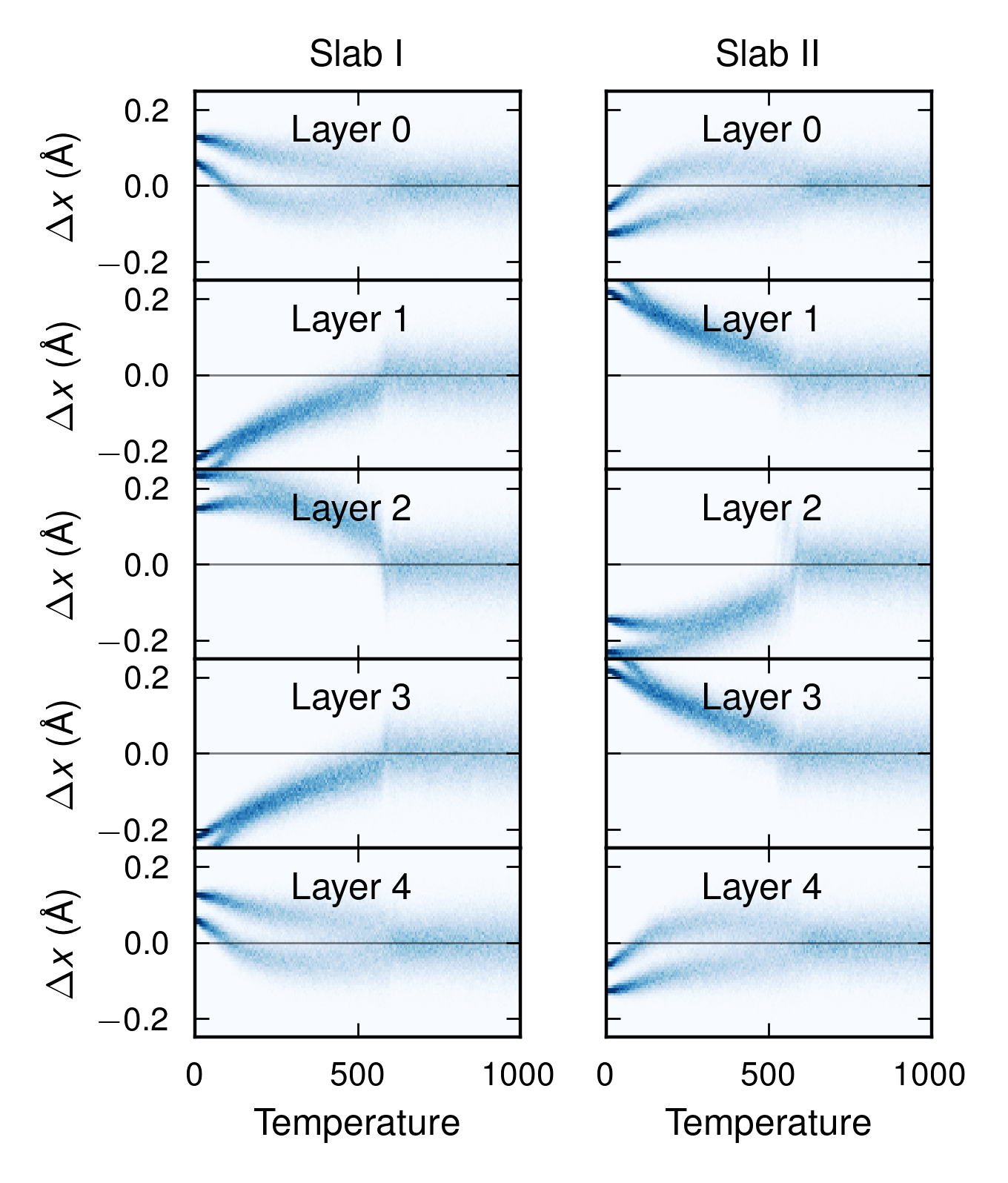}
    \includegraphics[width=0.49\linewidth]{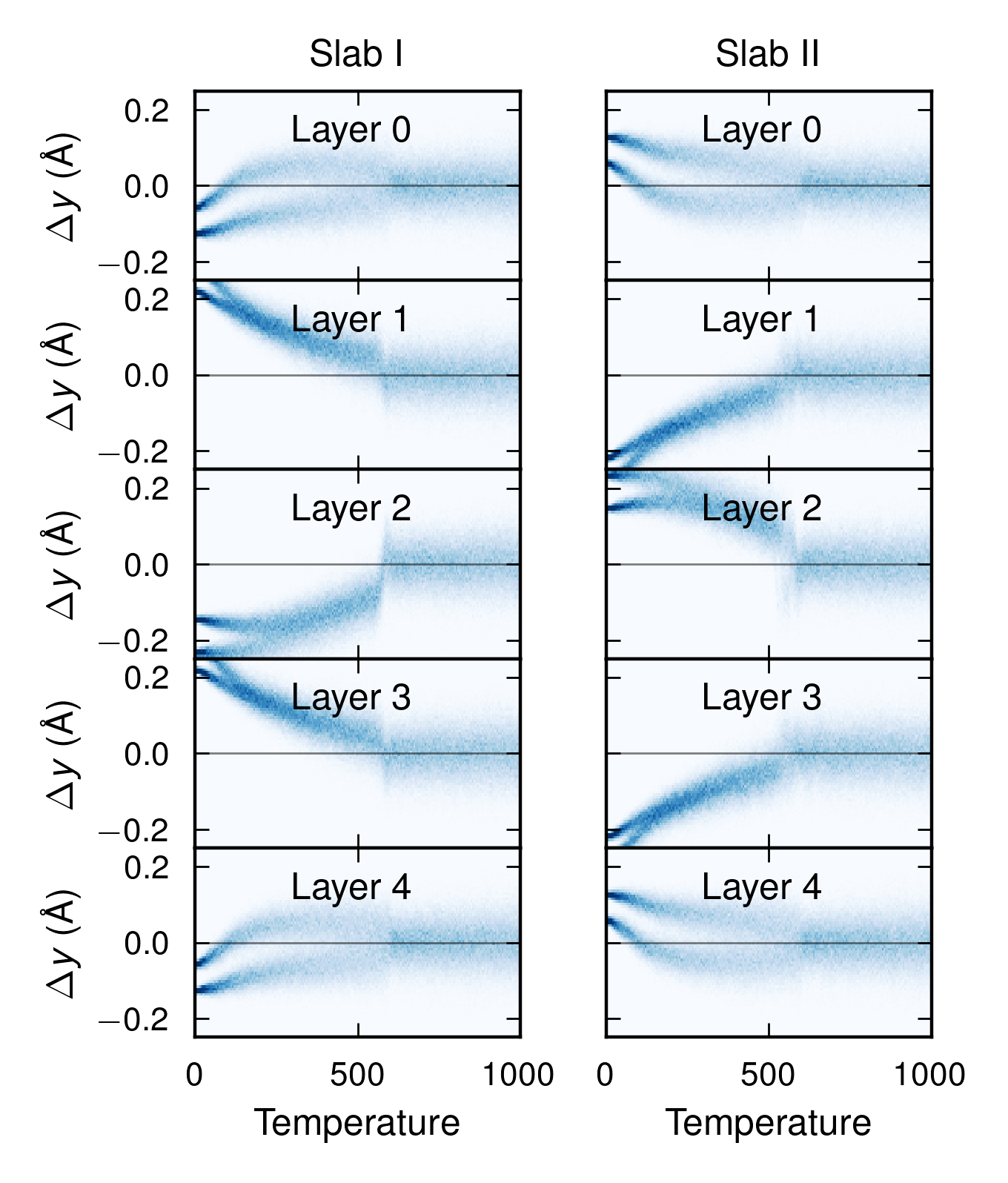}
    \caption{A-site displacements in the $n=4$ RP phase}
    \label{sfig:A_site_xy_n4}
\end{figure}

\clearpage
\section{X-ray diffraction}
X-ray diffraction (XRD) patterns were obtained with the \textsc{vesta} software using experimental values for the lattice parameters \cite{momma2011vesta}.

\begin{figure}[H]
    \centering
    \includegraphics[width=\linewidth]{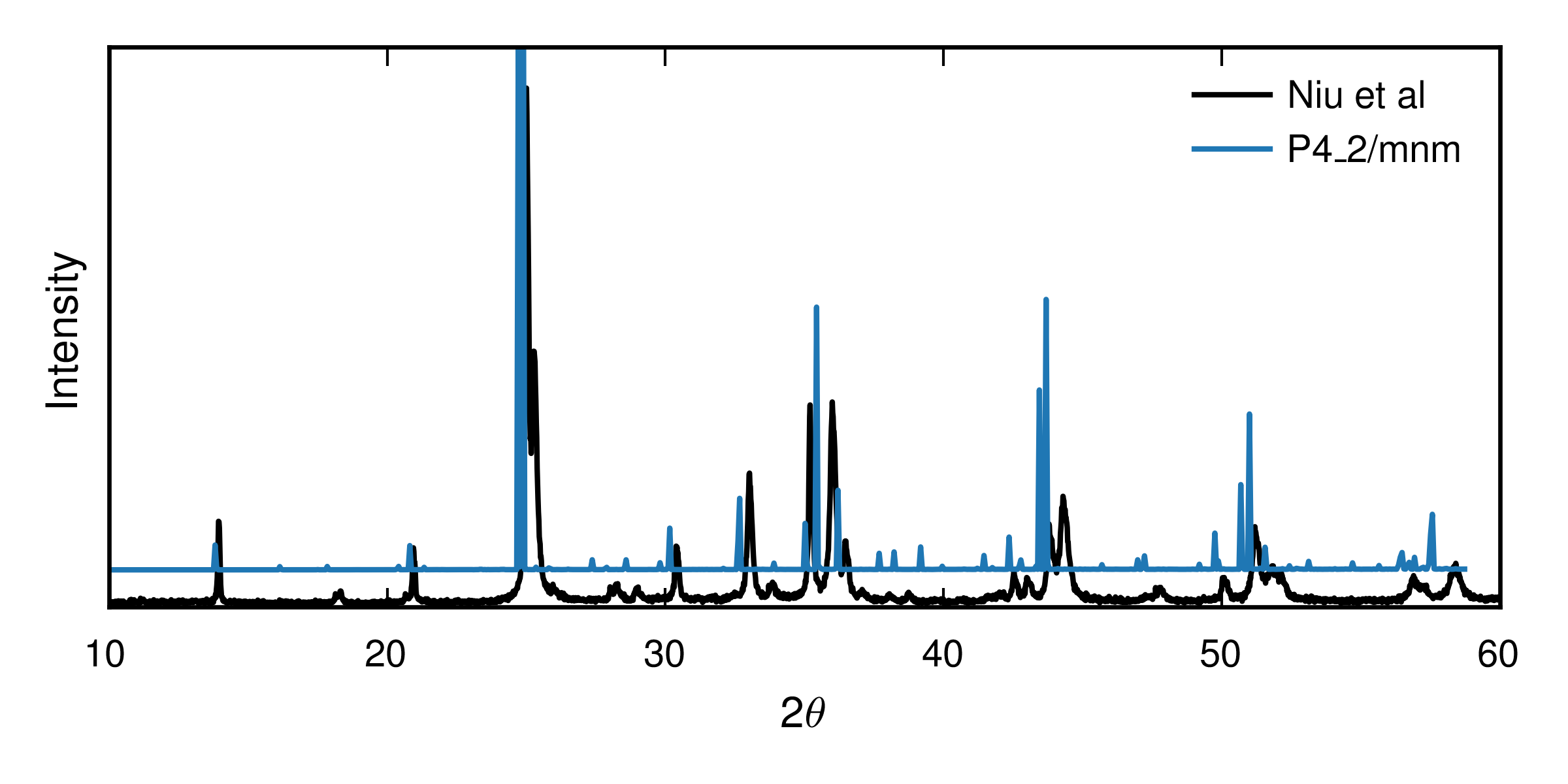}
    \caption{X-ray diffraction patterns for different phases of \ce{Ba3Zr2S7}. Experimental results are from Ref. \citenum{niu2019crystal}.}
    \label{fig:XRD_n2}
\end{figure}

In \autoref{fig:XRD_n2}, we show excellent agreement with the experimentally reported \hmn{P4_2/mnm} phase at room temperature. 

\begin{figure}[H]
    \centering
    \includegraphics[width=\linewidth]{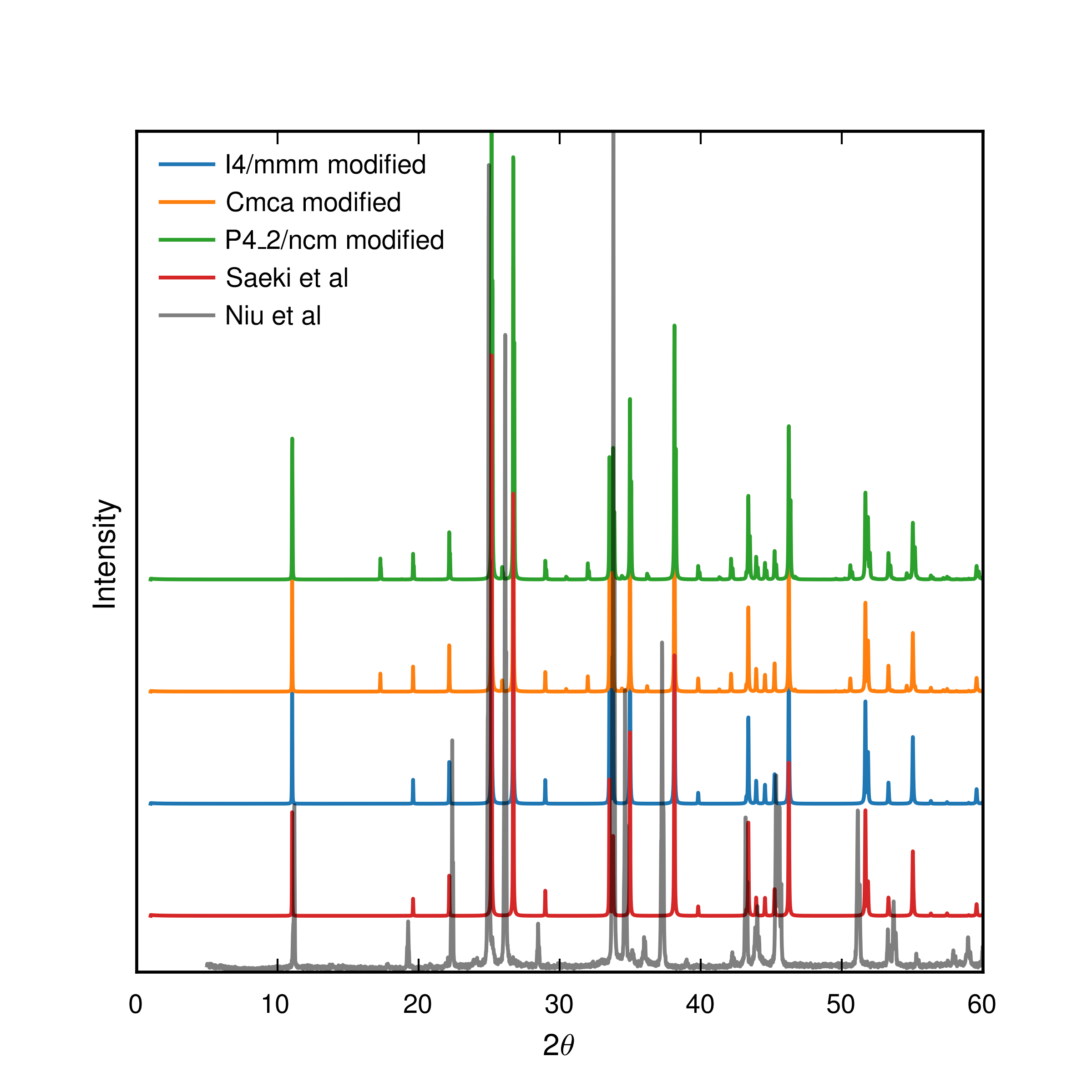}
    \caption{X-ray diffraction patterns for different phases of \ce{Ba2ZrS4}. Experimental results are from Ref. \citenum{saeki1991preparation} and Ref. \citenum{niu2018optimal}.}
    \label{fig:XRD_n1}
\end{figure}

In \autoref{fig:XRD_n1}, we show the simulated XRD for three different phases of the $n=1$ RP material -- the low temperature \hmn{P4_2/ncm} polymorph, the intermediate \hmn{Cmca}, and the high temperature \hmn{I4/mmm} polymorph. We also plot the experimentally reported data from Ref.\citenum{niu2019crystal} for comparison (where it is assigned to \hmn{I4/mmm}). 

We find that, as expected for such closely-related structures, the XRD patterns of all phases are highly similar.
At room temperature our heating simulations predict the \hmn{Cmca} phase to be stable. 
This assignment is supported by peak shoulders at \ang{25} and \ang{36} in the experimental XRD.
However, there are also a number of peaks missing in the experimental XRD at \ang{18}, \ang{32}, and \ang{50}.
XRD patterns can have broad and unresolved peaks for polycrystalline samples of \ce{BaZrS3} \cite{pacsca2025machine}. 
In addition, we note that the peaks associated with the \hmn{Cmca} will gradually decrease in intensity with increasing temperature due to its predicted second-order transition to the \hmn{I4/mmm} phase at \SI{400}{\kelvin}.
The phase transition temperatures are sensitive to the small energy differences between polymorphs, and so although we can tentatively assign the experimental XRD pattern to the \hmn{Cmca} phase, we cannot rule out formation of 
the \hmn{P4_2/ncm} or \hmn{I4/mmm} phases at room temperature. 
Further XRD measurements on a single crystal sample, combined with complementary techniques such as Raman spectroscopy, could aid in a more definite determination of the crystal structure. 

\clearpage
\phantomsection
\addcontentsline{toc}{section}{\listreferencename}


\bibliography{prl_reformatted.bib}